\documentclass[11pt]{amsart}
\usepackage{setspace}
\usepackage{amsmath,amstext,amssymb,amsthm}
\usepackage{eurosym}
\usepackage{graphicx}
\usepackage{mathrsfs}
\usepackage{xcolor}
\usepackage{hyperref}
\usepackage{enumerate}  
\usepackage{comment}
\allowdisplaybreaks
\usepackage{geometry}
\usepackage{multirow}
\usepackage{multicol}
\usepackage{listings}
\usepackage{breakcites}
\usepackage[font=footnotesize]{caption}
\numberwithin{equation}{section}

\newtheorem{theorem}{Theorem}[section]

\theoremstyle{definition}
\newtheorem{remark}[theorem]{Remark}

\title[A HMM for statistical arbitrage in international crude oil futures markets]{A hidden Markov model for statistical arbitrage \\ in international crude oil futures markets}

\author[V. Fanelli]{Viviana Fanelli}
\address{Department of Economics, University of Foggia, Italy. }
\email{viviana.fanelli@unifg.it}

\author[C. Fontana]{Claudio Fontana}
\address{Department of Mathematics ``Tullio Levi - Civita'', University of Padova, Italy.}
\email{fontana@math.unipd.it}

\author[F. Rotondi]{Francesco Rotondi}
\address{Department of Finance, Bocconi University, Italy.}
\email{rotondi.francesco@unibocconi.it}

\date{\today} 
\thanks{
	Financial support from the University of Padova (research programme STARS StG PRISMA - ``Probabilistic Methods for Information in Security Markets'') is gratefully acknowledged.
	This work is supported by the project P20224TM7Z (subject area: PE - Physical Sciences and Engineering) ``Probabilistic methods for energy transition'' funded by the European Union - Next Generation EU under the National Recovery and Resilience Plan, Mission 4 Component 1 CUP C53D23008390001.}

\begin{document}

\begin{abstract}
We study statistical arbitrage strategies in international crude oil futures markets. We analyse strategies that extend classical pairs trading strategies, considering two benchmark crude oil futures (Brent and  WTI) together with the recently introduced Shanghai crude oil futures. We show that the time series of these three futures prices are cointegrated and we introduce a mean-reverting regime-switching process modulated by a hidden Markov chain to model the cointegration spread. By relying on this model and applying online filter-based parameter estimators, we implement and test several statistical arbitrage strategies. Our analysis shows that: (i) arbitrage strategies involving the recently introduced Shanghai futures can be profitable even under conservative levels of transaction costs; (ii) strategies based on our model outperform those relying solely on observed spread values; and (iii) incorporating three futures contracts enables the implementation of arbitrage strategies even in cases where pairwise cointegration is not detected. 
\end{abstract}

\maketitle

\section{Introduction}

Pairs trading strategies represent a well-known instance of statistical arbitrage strategies that exploit temporary deviations in the prices of similar securities from their long-term equilibrium in order to achieve profits when convergence to the equilibrium is reached (see, e.g., \cite{Vidya}, \cite{EHM05}, \cite{GGR06}). Since one cannot determine ex ante when prices are going to realign, such strategies do not constitute pure arbitrage opportunities but rather statistical arbitrage opportunities, that are expected to deliver a profit over a sufficiently long time horizon. The profitability of this type of strategies requires a strong long-term relationship, so that mispricings are temporary and likely to revert back quickly. This is typically captured through the existence of a cointegration relation among the considered security prices. 

In this work, we study statistical arbitrage strategies in international crude oil futures markets, inspired by pairs trading strategies. The futures contracts we consider are two established benchmark crude oil futures, the Brent and the West Texas Intermediate (WTI), together with the  Shanghai crude oil futures more recently introduced in 2018\footnote{The Shanghai International Energy Exchange (INE) introduced crude oil futures on March 26, 2018. These contracts are known as the {\em Shanghai crude oil futures} and are traded on the Shanghai Futures Exchange (SHFE). They are quoted in Chinese yuan and based on the INE's own benchmark, which is a blend of various Middle Eastern and Asian crude oil grades that reflects the specific needs of China's oil imports.}. It is known that the Brent and the WTI futures prices are usually found to be cointegrated (see, e.g., \cite{KV14}, \cite{cerqueti2019long}, \cite{cerqueti2021long}, \cite{cotter2022commodity}). Nevertheless, the Shanghai crude oil futures contract exhibits institutional and structural differences that can contribute to temporary segmentation from these traditional benchmarks. These differences include its denomination in Chinese Renminbi, trading restrictions and a benchmark composition based on Middle Eastern and Asian crude grades. Recent studies (e.g., \cite{yang2020pricing}, \cite{HH20}) indicate that while the Shanghai market is gradually becoming more aligned with global benchmarks, it still shows signs of short-term divergences that present attractive opportunities for our statistical arbitrage strategies. These empirical findings motivate our joint cointegration analysis involving Brent, WTI, and Shanghai futures, and provide a compelling rationale for exploring potential statistical arbitrage opportunities in a globally integrated yet partially segmented crude oil futures market.

We examine whether the introduction of the Shanghai crude oil futures has enabled international arbitrageurs to achieve profitable investments through statistical arbitrage strategies. Pairs trading on traditional crude oil benchmarks has already been studied (see, e.g., \cite{DLE06}, \cite{CB12}, \cite{BB19}), while \cite{NMC2023} test pairs trading strategies on the different Shanghai futures contracts traded at the INE. To the best of our knowledge, our work is the first to consider statistical arbitrage strategies involving the recently introduced Shanghai crude oil futures.

From a methodological viewpoint, we improve on existing approaches to pairs trading by assessing the cointegration among the three futures prices at the same time and modelling the resulting cointegration spread by a mean-reverting process with regime-switching modulated by a hidden Markov chain. This generalizes the approach of \cite{TM17} and \cite{EB18} and enables us to capture time-varying cointegration regimes. As the Markov chain is unobserved, we employ stochastic filtering techniques to estimate the current regime and the model parameters, similarly as in \cite{EM09} and \cite{EBM10}. Parameter estimation is done by means of a filter-based version of the Expectation Maximization (EM) algorithm, as introduced in \cite{EAM} (see also \cite{FR10} for an application to credit risk). The filtering approach enables us to estimate the most likely regime and the model parameters  in a dynamic way, thereby ensuring that the model stays constantly tuned to the actual market situation. In turn, this will enable us to construct statistical arbitrage strategies that are dynamically updated as new information arrives. After estimating our model over a training sample, we move out-of-sample and analyse several types of statistical arbitrage strategies.

Our empirical analysis shows that statistical arbitrage strategies involving the three futures (Brent, WTI and Shanghai) at the same time deliver significant investment performances even when transaction costs are accounted for and even in cases where cointegration does not hold between any individual pair of these contracts.
Moreover, strategies that exploit our hidden Markov model for the cointegration spread are remarkably profitable when compared to more traditional strategies based only on time series features. 
Our findings indicate that the greater profitability of statistical arbitrage strategies involving the Shanghai futures can be explained by the higher speed of adjustment of the Shanghai futures prices compared to the Brent and the WTI. This implies that the Shanghai futures prices tend to revert back quickly to the long-run relationship, making easier the exploitation of temporary mispricings. This finding is in line with the fact that the profitability of statistical arbitrage strategies tends to decline over time as markets develop (see, e.g., \cite{DF10}), while the introduction of a new security can represent a valuable opportunity for arbitrageurs\footnote{See for instance \cite{CrepellierePelsterZeisberger2023} for an analysis of arbitrage opportunities in the emerging markets for cryptocurrencies.}.

The contribution of the paper is threefold. First, we corroborate the evidence of cointegration among the recently introduced Shanghai crude oil futures and two long-standing crude oil futures benchmarks, the Brent and the WTI. To our knowledge, this is the first time that cointegration is assessed considering these three contracts together, rather than on a pairwise basis. As a second contribution, we propose a mean-reverting stochastic model for the cointegration spread that allows for regime switching by means of a hidden Markov chain determining the parameters of the process. We apply filtering techniques to dynamically estimate the most likely regime and the model parameters. Finally, we empirically analyze different statistical arbitrage strategies involving three crude oil futures contracts. Our results indicate that strategies that include a newly introduced security, like the Shanghai crude oil futures, along with traditional and well-established ones, like the Brent and the WTI, are remarkably profitable and robust.

The paper is structured as follows. Section \ref{sec:Cointegration} describes the cointegration structure and the stochastic model for the cointegration spread. Section \ref{sec:StatArbStrategies} describes the statistical arbitrage strategies we implement and how their performance is assessed. Section \ref{sec:Empirics} contains the results of our empirical analysis, while 
Section \ref{sec:Conclusions} concludes. 
The results of an additional empirical analysis based on daily data are reported as Supplementary Material.

\section{Cointegration analysis and spread modelling}\label{sec:Cointegration}

In this section, we introduce the stochastic model for the long-run relationship among the three crude oil future prices (Section \ref{subsec:OUHMM}) and describe how the model parameters can be dynamically estimated by means of filtering techniques combined with the EM algorithm (Section \ref{subsec:filter_EM}).

\subsection{A mean-reverting hidden Markov model}\label{subsec:OUHMM}

Let $F^i=(F^i_t)_{t\in\mathbb{N}}$ denote the futures price process\footnote{We consider statistical arbitrage strategies based on the price processes of several assets. We refer to \cite{FocardiFabozziMitov2016} for an analysis of statistical arbitrage opportunities based on asset returns rather than prices.}, for $i\in\{B,S,W\}$, of the Brent ($B$), the Shanghai ($S$), and the WTI futures ($W$).
The processes $F^B$ and $F^W$ are typically found to be cointegrated, as documented by 
%\cite{LMG05}
\cite{HET08} and \cite{KV14} among others. Similarly,  \cite{YZ20} and \cite{HH20} show that $F^S$ is cointegrated with $F^B$ and with $F^W$. Therefore, it is reasonable to guess that the three futures price processes jointly considered are cointegrated, similarly to the case of the Brent, the Dubai and the WTI futures prices (see \cite{G19}). The cointegration among $F^B$, $F^S$, $F^W$ has not been tested in previous works and will be shown to hold in the following.
In the presence of cointegration, there exists a linear combination of $F^B$, $F^S$, $F^W$ that follows a stationary and possibly mean-reverting process. As usual in cointegration analysis, we call this linear combination the \emph{spread process} and denote it by $S = (S_t)_{t \in\mathbb{N}}$:
\begin{equation}\label{eqn:spreadWeights}
    S_t := \lambda^0 + \sum_{i = B,S,W} \lambda^i F_t^i,
    \qquad\text{ for all }t\in\mathbb{N},
\end{equation}
for suitable coefficients $\lambda^0, \lambda^B,\lambda^S,\lambda^W$ that constitute the \emph{cointegration vector} $\lambda$. 

Using the cointegration vector estimated in-sample, an investor can evaluate  the spread out-of-sample and set up statistical arbitrage strategies based on it, taking as portfolio weights the elements of the cointegration vector (see Section \ref{sec:StatArbStrategies} for more details). 
These strategies turn out to be profitable as long as the cointegrating relationship is stable, namely as long as the spread $S$ is stationary. 
In a standard cointegration analysis, the cointegration vector $\lambda$ is assumed to be constant over time. However, previous works on the Brent and on the WTI document the presence of structural breaks in the cointegration vector (see, e.g., \cite{CFT19}), showing also that the cointegration vector depends significantly on the  time window under consideration.
As already noted by \cite{LP16}, this might severely impact the performance of pairs trading strategies based on cointegration. 
It is therefore appropriate to model the spread as a stochastic process with regime switching, updating dynamically the model parameters as new information becomes available. In this work, we adopt this approach and explicitly model the spread as a regime-switching stochastic process.

\textcolor{black}{Let $(\Omega, \mathcal{F}, \mathbb{F} = (\mathcal{F}_t)_{t \in\mathbb{N}}, \mathbb{P})$ be a filtered probability space supporting a sequence $(z_t)_{t\in\mathbb{N}}$ of i.i.d. standard normal random variables and a Markov chain $\textbf{X} = (\textbf{X}_t)_{t \in\mathbb{N}}$ with $N$ states and transition matrix $\pmb{\Pi}$. 
The states of the Markov chain $\textbf{X}$ represent different market regimes, thus capturing the time-varying nature of the cointegrating relationship.
The Markov chain $\textbf{X}$ admits a representation of the form
\[
    \textbf{X}_{t+1} = \pmb{\Pi} \textbf{X}_t + \textbf{M}_{t+1},
    \qquad\text{ for all }t\in\mathbb{N},
\]
where $\textbf{X}_0$ is the initial state of the Markov chain and $\textbf{M}=(\textbf{M}_t)_{t\in\mathbb{N}}$ is a sequence of martingale increments with $\textbf{M}_0=0$ (see \cite{EAM}, Chapter 2).} 

\textcolor{black}{For the spread process $S$, we assume autoregressive dynamics with regime switching:
\begin{equation}\label{eqn:discreteTimeSpread}
    S_{t+1} = \gamma (\textbf{X}_t) + \alpha(\textbf{X}_t) S_t + \eta (\textbf{X}_t) z_t,
    \qquad S_0 = s_0 \in \mathbb{R},
\end{equation}
where $\gamma$, $\alpha$ and $\eta$ are real-valued functions on $\mathbb{R}^N$.
Without loss of generality and similarly as in \cite{EAM}, we assume that the state space of the Markov chain $\textbf{X}$ is  the canonical basis of $\mathbb{R}^N$, denoted by $\{ \textbf{e}_1 ,\ldots, \textbf{e}_N \}$, so that $\gamma(\textbf{X}_t) = \pmb{\gamma}^\top \textbf{X}_t$ with $\pmb{\gamma} \in \mathbb{R}^N$. 
The term $\gamma_i$ represents the intercept of $S$ in the $i^{\text{th}}$ regime, while $\alpha_i$ and $\eta_i$ represent respectively the autoregressive coefficient and the volatility of $S$ in the $i^{\text{th}}$ regime, for each $i=1, \dots, N$.
The condition $|\alpha_i|<1$, ensures that the spread is mean-reverting within each regime and admits a unique stationary distribution conditionally on that regime.
The resulting model for $S$ is an autoregressive process of order one modulated by a Markov chain, widely used in the time-series literature to capture nonlinear dynamics and regime-dependent persistence (see, e.g., \cite{Hamilton}, Chapter 22), which can effectively affect the spread process in our context.}

Since the current market regime is generally unknown by market participants, we assume that the Markov chain $\textbf{X}$ is unobservable, giving rise to an autoregressive process modulated by a hidden Markov model (AR-HMM). As a consequence, the current state of $\textbf{X}$ has to be filtered from the observations of the spread. 
Since we want to implement statistical arbitrage strategies that exploit the probabilistic structure of the model (\ref{eqn:discreteTimeSpread}) for $S$, we need to estimate $\pmb{\gamma}$, $\pmb{\alpha}$, $\pmb{\eta}$ and the transition matrix $\pmb{\Pi}$. In this partial observation setup, the estimation of $\pmb{\gamma}$, $\pmb{\alpha}$, $\pmb{\eta}$ and $\pmb{\Pi}$ can be done by a filter-based Expectation Maximization algorithm that we describe below.

\begin{remark}\label{rem:continuous_time}
\textcolor{black}{
The regime-switching autoregressive model introduced above can be seen as a discrete-time representation, at the observation frequency, of a continuous-time Ornstein-Uhlenbeck process modulated by a hidden Markov model (OU-HMM).
Specifically, let $\textbf{X} = (\textbf{X}_t)_{t \geq 0}$ be a continuous-time Markov chain. The continuous-time analogue of the discrete-time spread dynamics \eqref{eqn:discreteTimeSpread} can then be written as follows:
\begin{equation}\label{eqn:continuousTimeSpread}
    \mathrm{d} S_t = a(\textbf{X}_t) \left( \beta (\textbf{X}_t) - S_t \right) \mathrm{d} t + \xi (\textbf{X}_t) \mathrm{d} W_t,
\end{equation}
where $W = (W_t)_{t \geq 0}$ is a Brownian motion. The functions $a$, $\beta$ and $\xi$ represent respectively the speed of mean-reversion, the long-run mean and the volatility of the spread as functions of the current regime. 
OU-HMMs have been used for the modelling of interest rates (\cite{EM09}, \cite{GEM20}), commodity spot prices (\cite{EBM10}) and spreads between equity price processes (\cite{EB18}).
The coefficients of \eqref{eqn:continuousTimeSpread} can be mapped into the coefficients of \eqref{eqn:discreteTimeSpread} as follows, where $\Delta$ denotes the observation frequency:
\[
 \alpha(\textbf{X}_t)= e^{ - a(\textbf{X}_t) \Delta },
 \qquad
\gamma(\textbf{X}_t) = \beta(\textbf{X}_t) \bigl( 1 - e^{ - a(\textbf{X}_t) \Delta } \bigr), 
\qquad        
\eta(\textbf{X}_t) = \xi (\textbf{X}_t) \sqrt{\frac{1-e^{ - 2a(\textbf{X}_t) \Delta }}{2a(\textbf{X}_t)}}.
\]
}

While continuous-time mean-reverting dynamics have been considered in several works on pairs trading (see for instance \cite{AL10,Ber10,XH22}), we emphasize that adopting the discrete-time formulation in \eqref{eqn:discreteTimeSpread} is not only motivated by the fact that observations are available at discrete intervals, but is also essential from an estimation perspective. In continuous time, the EM algorithm relies on equivalent measure transformations, under which volatility parameters are invariant. As a consequence, the volatility function cannot be estimated by a continuous-time EM algorithm, whereas the discrete-time formulation allows all model parameters, including volatility, to be consistently estimated.
\end{remark}

\subsection{The filter-based Expectation Maximization algorithm}\label{subsec:filter_EM}

We now describe the filter-based Expectation Maximization (EM) algorithm for the estimation of the AR-HMM in (\ref{eqn:discreteTimeSpread}). The following results are adapted from the general techniques described in \cite{EAM} and more specifically in \cite{EM09} for HMM-modulated autoregressive processes with Gaussian innovations.
As a preliminary, we introduce the following auxiliary quantities that will be needed for the recursive estimation of the model parameters:
\begin{itemize}
    \item $J^{ij}_{t} := \sum_{n=1}^t \langle \mathbf{X}_{n-1}, \mathbf{e}_i \rangle \langle \mathbf{X}_n , \mathbf{e}_j \rangle$, representing the cumulative number of jumps of $\mathbf{X}$ from state $\mathbf{e}_i$ to state $\mathbf{e}_j$ until time $t$. 
    % Notice that $J^{ij}_{t} = J^{ij}_{t-1} + \langle \mathbf{X}_{t-1}, \mathbf{e}_i \rangle \langle \mathbf{X}_t , \mathbf{e}_j \rangle$.
    \item $O^i_t := \sum_{n=1}^t \langle \mathbf{X}_n, \mathbf{e}_i \rangle$, representing the occupation time of $\mathbf{X}$ in state $\mathbf{e}_i$ until time $t$. 
    % Notice that $O^i_t = O^i_{t-1} + \langle \mathbf{X}_t, \mathbf{e}_i \rangle$.
    \item $T^i_t (f) = \sum_{n=1}^t \langle \mathbf{X}_{n-1}, \mathbf{e}_i \rangle f_n $, where $f_n$ is a generic function of the observations of $S$ up to time $n$ (in our case, the function $f_n$ will be given by $f_n = S_n$, $f_n = S_n^2$ or $f_n = S_n S_{n-1}$). 
\end{itemize}

We denote by $\mathbb{F}^S = (\mathcal{F}_t^S)_{t \in \mathbb{N}}$, with $\mathcal{F}^S_t := \sigma \left\{ S_0, \dots, S_t \right\}$ for all $t\in\mathbb{N}$,  the filtration generated by the spread process $S$. 
We assume that $\mathbb{F}^S$ represents the information available to the investor, who cannot  observe the Markov chain $\textbf{X}$ and can only access current and past discrete-time observations of the spread.

We now state the recursive filtering equations for the unobserved Markov chain $\mathbf{X}$ and the quantities $J$, $O$, $T$, from which we derive  the recursive equations for the EM estimators of $\pmb{\alpha}$, $\pmb{\gamma}$, $\pmb{\eta}$ and $\pmb{\Pi}$. 
Those filtered estimates will enable the investor to compute one-step-ahead forecasts of the conditional mean and variance of the spread, which will be used to design model-based statistical arbitrage strategies in Section \ref{sec:StatArbStrategies}.
For $t\in\mathbb{N}$, we denote by $\hat{\mathbf{X}}_t := \mathbb{E} \left[ \mathbf{X}_t | \mathcal{F}^S_t \right]$  the filtered estimate of the latent Markov chain at time $t$ on the basis of the information generated by the observation process $S$ up to time $t$. In an analogous way we define the filtered estimates $\hat{J}_{t}^{ij}$, $\hat{O}_{t}^{i}$ and $\hat{T}_{t}^{i}(f)$ of the quantities introduced above. 
Moreover, we denote by $\hat{\pmb{\alpha}}^{(t)} = (\hat{\alpha}_i^{(t)})_{i = 1, \dots, N}$ the estimate of the parameter $\pmb{\alpha}$ on the basis of the information generated by the observation process $S$ up to time $t$. The quantities  $\hat{\pmb{\gamma}}^{(t)}$, $\hat{\pmb{\eta}}^{(t)}$, $\hat{\pmb{\Pi}}^{(t)}$ are defined analogously.

As explained in \cite[Chapter~2]{EAM}, in order to filter the unobserved Markov chain $\mathbf{X}$ one can resort to a change of measure passing from $\mathbb{P}$ to an equivalent probability measure $\tilde{\mathbb{P}}$ under which the observation process $S$ is independent of the Markov chain $\mathbf{X}$. For each $T\in\mathbb{N}$, the probability measures $\mathbb{P}$ are $\tilde{\mathbb{P}}$ are related by the Radon-Nikodym derivative 
$$ \Lambda_T = \left. \frac{\mathrm{d} \mathbb{P}}{ \mathrm{d} \tilde{\mathbb{P}}} \right|_{\mathcal{F}_T} = \prod_{t = 1}^T \lambda_t, $$
where
\begin{equation}\label{eqn:lambda_t}
    \lambda_t =  \exp \left( -\frac{1}{\eta( \mathbf{X}_{t-1})^2} \left( \left( \alpha( \mathbf{X}_{t-1}) S_{t-1} +\gamma( \mathbf{X}_{t-1}) \right)S_t + \frac{1}{2} \left( \alpha( \mathbf{X}_{t-1}) S_{t-1} +\gamma( \mathbf{X}_{t-1}) \right)^2 \right) \right).
\end{equation}
with $\Lambda_0 = 1$.
For convenience of notation, it is useful to represent the possible values of $\lambda_t$ in (\ref{eqn:lambda_t}) associated to each of the $N$ states of $\mathbf{X}$ by a diagonal matrix $\mathbf{D}_t = [ d^{ij}_t ]_{i,j=1,...,N}$ with
\begin{equation}\label{eqn_apx:ds}
    d^{ii}_t =  \exp \left( -\frac{1}{\eta_i^2} \left( \left( \alpha_i S_{t-1} +\gamma_i \right)S_t + \frac{1}{2} \left( \alpha_i S_{t-1} +\gamma_i \right)^2 \right) \right),
    \qquad\text{ for }i=1,\ldots,N.
\end{equation}

The starting point of the filtering algorithm is a set of initial guesses for the quantities $\mathbf{X}_0$ (that delivers also the initial values of $O^i_{0}$) and for $\pmb{\Pi}$, $\pmb{\gamma}, \pmb{\alpha}$, $\pmb{\eta}$, see Remark \ref{remark:implementationFilters} for more details.

By performing similar computations as in \cite{EM09}, the recursive filtering equations for $\mathbf{X}$ and for the auxiliary quantities $J$, $O$, $T$ are given as follows:
\begin{equation}\label{eqn_apx:filters}
    \begin{array}{rl}
        \hat{\mathbf{X}}_{t+1} &= \pmb{\Pi} \mathbf{D}_{t+1} \hat{\mathbf{X}}_t, \\
        \hat{J}^{i,j}_{t+1} &= \langle \pmb{1}, \pmb{\Pi} \mathbf{D}_{t+1} \hat{J}^{i,j}_t + \langle \hat{\mathbf{X}}_{t}, \mathbf{e}_i \rangle \langle \mathbf{D}_{t+1} \mathbf{e}_i, \mathbf{e}_i \rangle \pi_{ij} \mathbf{e}_j  \rangle, \\
        \hat{O}^i_{t+1} &= \langle \pmb{1}, \pmb{\Pi} \mathbf{D}_{t+1} \hat{O}^{i}_t + \langle \hat{\mathbf{X}}_t, \mathbf{e}_i \rangle \langle \mathbf{D}_{t+1} \mathbf{e}_i, \mathbf{e}_i \rangle \pmb{\Pi} \mathbf{e}_i \rangle, \\
        \hat{T}_{t+1}^i(f) &= \langle \pmb{1}, \pmb{\Pi} \mathbf{D}_{t+1} \hat{T}_t^i(f) + \langle \hat{\mathbf{X}}_t, \mathbf{e}_i \rangle \langle \mathbf{D}_{t+1} \mathbf{e}_i, \mathbf{e}_i \rangle f_t \pmb{\Pi} \mathbf{e}_i \rangle,
    \end{array}
\end{equation}
for all $t \in \mathbb{N}$, where we denote by $\pmb{1}$ the unit vector in $\mathbb{R}^N$. 
Observe that the filters in \eqref{eqn_apx:filters} depend on $\pmb{\Pi}$ and, through the matrix $\mathbf{D}_{t+1}$, also on $\pmb{\gamma}$, $\pmb{\alpha}$, $\pmb{\eta}$. Therefore, when implementing the filtering procedure in practice these quantities must be replaced by their estimates $\hat{\pmb{\Pi}}^{(t)}$, $\hat{\pmb{\gamma}}^{(t)}$, $\hat{\pmb{\alpha}}^{(t)}$, $\hat{\pmb{\eta}}^{(t)}$ obtained at the previous time point $t$.
In analogy to \cite{EM09}, the equations for the EM recursive estimates of $\pmb{\Pi}$, $\pmb{\gamma}, \pmb{\alpha}$, $\pmb{\eta}$ at time $t$ are given by
\begin{equation}\label{eqn_apx:estimationEqs}
    \begin{array}{rl}
        \hat{\pi}_{ij}^{(t+1)} &= \displaystyle\frac{\hat{J}^{i,j}_{t+1}}{\hat{O}^j_{t+1}}, \\
        \hat{\gamma}_i^{(t+1)} &= \displaystyle\frac{\hat{T}_{t+1}^i(S_{t+1}) - \hat{\alpha}_i^{(t)} \hat{T}_{t}^i(S_t) }{\hat{O}^i_{t+1}}, \\
        \hat{\alpha}_i^{(t+1)} &= \displaystyle\frac{\hat{T}_{t}^i(S_{t+1},S_{t}) - \hat{\gamma}_i^{(t+1)} \hat{T}_{t}^i(S_t) }{\hat{T}_{t}^i(S_{t}^2)}, \\
        \bigl( \hat{\eta}_i^{(t+1)} \bigr)^2 &= \displaystyle\frac{\hat{T}_{t+1}^i(S_{t+1}^2) + (\hat{\alpha}_i^{(t+1)})^2 \hat{T}_{t}^i(S_{t}^2) +(\hat{\gamma}_i^{(t+1)})^2 \hat{O}^i_{t+1}}{\hat{O}^i_{t+1}} \\
        & \quad - \displaystyle\frac{2\hat{\gamma}_i^{(t+1)} \hat{T}_{t+1}^i(S_{t+1}) +2  \hat{\alpha}_i^{(t+1)} \hat{T}_{t}^i(S_{t+1},S_{t}) +2\hat{\alpha}_i^{(t+1)} \hat{\gamma}_i^{(t+1)} \hat{T}_{t}^i(S_t) }{\hat{O}^i_{t+1}}. \\
    \end{array}
\end{equation}
It is worth noting that every time a new observation becomes available all the filters in (\ref{eqn_apx:filters}) and (\ref{eqn_apx:estimationEqs}) are simply updated without re-running the whole filtering algorithm, due to its recursive nature. This is an advantage of the online filter-based EM algorithm, which allows the model to be constantly tuned to the actual market information in a computationally efficient way.

\begin{remark}[Numerical aspects]\label{remark:implementationFilters}
In this remark we address some numerical aspects that have to be taken into account when implementing the above filtering algorithm in practice: 
\begin{enumerate}
    \item Concerning the initialization of the algorithm,  the initial state of the Markov chain $\mathbf{X}_0$ also determines $O^i_{0}$. In our analysis we set $\hat{\mathbf{X}}_0 = \mathbf{e}_1$ by convention. This implies $\hat{O}^1_{0} = 1$ and $\hat{O}^i_{0} = 0$, for $i=2, \dots, N$. With respect to the initial guess for $\pmb{\Pi}$, if $N=2$, we set $\hat{\pi}_{11}^{(0)} = 0.6$ and $\hat{\pi}_{22}^{(0)} = 0.5$ since the perfectly symmetric initial guess $\hat{\pi}_{11}^{(0)} = \hat{\pi}_{22}^{(0)} = 0.5$ tends to create numerical instabilities as there is no difference across states in the transition matrix. If $N=3$, we set $\hat{\pi}_{11}^{(0)} = 0.5, \hat{\pi}_{12}^{(0)} = 0.25$, $\hat{\pi}_{21}^{(0)} = 0.3, \hat{\pi}_{22}^{(0)} = 0.4$ and $\hat{\pi}_{31}^{(0)} = 0.2, \hat{\pi}_{33}^{(0)} = 0.6$. \item The initial guesses for the parameters $\pmb{\gamma}, \pmb{\alpha}$, $\pmb{\eta}$ are derived from the data. More specifically, in the absence of regime switching, the spread process would follow a simple AR(1) process with constant parameters satisfying
    \begin{equation}\label{eqn:LinearModel}
        S_{t+1} = \gamma + \alpha S_t + \varepsilon_{t+1},   
        \qquad\text{ for all }t\in\mathbb{N},
    \end{equation}
    where $\varepsilon = (\varepsilon_t)_{t \in \mathbb{N}}$ is a Gaussian white noise process with variance $\eta^2$. In this case, the parameters $\gamma$, $\alpha$, $\eta$ can be estimated by ordinary least squares (OLS). In our empirical analysis we do so using the first $20$ datapoints of $S$, corresponding to one month of daily observations (twice the batch parameter $m$ introduced below). 
    For $N=1$, we set $\hat{\gamma}^{(0)} = \hat{\gamma}_{\rm OLS}$, $\hat{\alpha}^{(0)} = \hat{\alpha}_{\rm OLS}$ and $\hat{\eta}^{(0)} = \hat{\eta}_{\rm OLS}$, where $\hat{\gamma}_{\rm OLS}$, $\hat{\alpha}_{\rm OLS}$ and $\hat{\eta}_{\rm OLS}$ denote the OLS estimates of the parameters of (\ref{eqn:LinearModel}). For $N=2$, we let the initial guesses of all parameters for the two states be equally spaced with respect to the OLS estimates determined as above. Hence, we set $\hat{\gamma}_1^{(0)} = 1.3 \hat{\gamma}_{\rm OLS}$ and $\hat{\gamma}_2^{(0)} = 0.7 \hat{\gamma}_{\rm OLS}$, and analogously for the other parameters. For $N=3$, we follow the same reasoning and set $\hat{\gamma}_1^{(0)} = 1.3 \hat{\gamma}_{\rm OLS}$, $\hat{\gamma}_2^{(0)} = \hat{\gamma}_{\rm OLS}$ and $\hat{\gamma}_3^{(0)} = 0.7 \hat{\gamma}_{\rm OLS}$, and similarly for the remaining  parameters. 
    \item \textcolor{black}{Due to the recursive and online nature of the filter-based EM algorithm, the specification of initial parameter values constitutes the only non-data-driven input of the algorithm. This aspect is particularly relevant in regime-switching models, whose likelihood function is generally non-concave and may admit multiple local maxima (see, e.g., \cite[Chapter 3]{McLachlanKrishnan}). To assess the stability of our results with respect to the initialization and address potential concerns related to convergence to local optima, we perform a systematic set of robustness checks by varying the initial values of all regime-dependent parameters and transition probabilities. As documented in Remarks \ref{remark:stabilityEstimates} and \ref{remark:stabilityPerformances}, while the resulting point estimates exhibit mild sensitivity to different initializations, the performance of the strategies derived from the model remains stable. This indicates that the results reported in Section \ref{sec:Empirics} are robust with respect to the initialization and are not affected by convergence to a particular local maximum of the likelihood.}
    \item During the first iterations of the  algorithm it might happen that the quantity $\hat{O}^i_{t+1}$, which appears at the denominator in $\hat{\gamma}_i^{(t+1)}$ and $\hat{\eta}_i^{(t+1)}$ in (\ref{eqn_apx:estimationEqs}), is equal to zero. This happens if the Markov chain has never visited state $i$ before time $t+1$. In this  case, there is no way to update $\hat{\gamma}_i^{(t+1)}$ and $\hat{\eta}_i^{(t+1)}$, which are left unchanged throughout the current iteration of the algorithm.
    \item Quantities like $\hat{T}_t^i(S_{t-1}^2)$, appearing at the denominator in $\hat{\alpha}_i^{(t+1)}$ and $\hat{\eta}_i^{(t+1)}$ in (\ref{eqn_apx:estimationEqs}), might take very small values (especially when the $S$'s are close to zero) and this might induce numerical instabilities. The same issue can arise when computing the diagonal elements of $\mathbf{D}$ if the $\hat{\eta}_i^{(t+1)}$'s are too close to zero. If not controlled for, these instabilities propagate, preventing the algorithm to converge. Therefore, for all denominators in (\ref{eqn_apx:estimationEqs}) we check that the new estimate of the quantity of interest is not smaller/larger than ten times the previous estimate and, in case, we truncate the new estimate to that level.
    \item The two previous issues can be mitigated by adopting a \emph{batchwise} approach, as pointed out in \cite{GEM20}, updating the parameters in (\ref{eqn_apx:estimationEqs}) not at every time point $t$ as the filters in (\ref{eqn_apx:filters}) but rather every $m$ steps. This technique stabilizes the estimates of the filtered quantities in (\ref{eqn_apx:filters}) and, as a consequence, the parameter estimates in (\ref{eqn_apx:estimationEqs}). In our analysis we set $m=10$, which amounts to updating the model parameters every two weeks.
\end{enumerate}   
\end{remark}

\section{Statistical arbitrage strategies}\label{sec:StatArbStrategies}

In this section we introduce several statistical arbitrage strategies that shall be tested (Section \ref{subsec:ParisTradingStrategies}) and describe the performance measures used to assess their profitability (Section \ref{subsec:PerformanceMeasure}).

\subsection{Construction of statistical arbitrage strategies}\label{subsec:ParisTradingStrategies}
As explained in \cite[Chapter~8]{Vidya}, the basic intuition behind pairs trading is that whenever the spread deviates ``sufficiently'' from its equilibrium value, the investor should open a trading position, appropriately investing in the underlying assets. The position is then closed when this deviation corrects itself and the spread reverts to its equilibrium value. The precise specification of what we mean above by ``sufficiently'' determines a specific trading strategy, which can be summarized by a simple rule based on \emph{opening} and \emph{closing} signals, as explained in the following.

\textcolor{black}{We consider self-financing trading strategies investing in futures contracts. Any cash not engaged in futures positions is implicitly invested in a riskless money market account, earning zero interest rate for simplicity of presentation. When no trading signal is active, the strategy holds no futures positions and therefore generates zero returns.}
Regardless of the specification of opening/closing signals, we assume that the trading position is determined by the cointegration vector $\lambda=(\lambda^0,\lambda^B,\lambda^S,\lambda^W)$ in \eqref{eqn:spreadWeights}, where $\lambda^0$ represents the cash component.
%determines the borrowing/lending of money from a riskless money market account (earning zero interest rate, for simplicity of presentation). 
Abstracting from transaction costs, if an opening signal is received at time $t$ when $S_t>0$ (resp.\ $S_t <0$), the investor has to go short (resp.\ long) on portfolio $\lambda$. This generates an inflow of money at time $t$ equal to $|S_t|$. A position opened at time $t-1$ is closed at time $t$ if the spread $S_t$ changes sign. 
\textcolor{black}{The cointegration vector $\lambda$ determines the relative composition of the portfolio. Since cointegration vectors are defined only up to scale, a normalization is required to define the trading strategies in a a meaningful and economically interpretable way. In our framework, the scale of the investment at time $t$ is set by normalizing the portfolio $\lambda$ by the gross notional exposure, defined as $G_t := |\lambda_0| + \sum_{k=B,S,W}  |\lambda_k|F^k_t$, for $t\in\mathbb{N}$. This quantity represents the total market value of the long and short futures positions implied by the cointegration vector at prevailing prices. This normalization will yield returns per unit of gross exposure and ensure that performance measures are invariant to rescaling the cointegration vector (see Section \ref{subsec:PerformanceMeasure}).}

The above trading rule corresponds to the first crossing of the spread through zero in discrete-time, consistent with standard practice in pairs trading. Closing the position as soon as the spread crosses zero will deliver either no cashflow or a small positive one. Therefore, these strategies are designed to deliver positive payoffs at their opening, with essentially zero cashflows when they are closed, similarly to arbitrage opportunities of the second kind (see, e.g., \cite{Ingersoll}). However, since it is not known ex ante if and when $S_t$ will revert back to zero, these strategies are not pure arbitrage opportunities but only statistical arbitrage opportunities (see \cite{Bon03} and \cite{RRS2021} for a formalization of statistical arbitrage).

\begin{remark}
We emphasize that, by design, our strategies are not zero-net investment strategies, which are a well-established approach in pairs trading. Instead, they can be described as ``negative-cost'' investment strategies: they generate a positive cash inflow at initiation and require no cash outflow at closure. Nevertheless, like traditional zero-net investment strategies, they do not require initial capital and are, in principle, scalable. Importantly, consistent with this feature, the returns reported in Subsection \ref{subsec:PerformanceMeasure} should be interpreted as excess returns.
\end{remark}

In our analysis we consider five different statistical arbitrage strategies. Strategy 1 serves as a simple benchmark and requires no explicit modelling of the spread process. Strategies 2 and 4 are backward-looking in nature and are based on sample estimates of the moments of the spread. By contrast, Strategies 3 and 5 exploit the stochastic model introduced in Section 2, as they are based on one-step-ahead forecast of the spread and on the corresponding forecast interval.

\textcolor{black}{
With the exception of the benchmark strategy, all strategies are designed to time both the opening and the closing of trades. These timing decisions are driven by crossings of the observed spread process with a set of bands, defined either as probability intervals or as forecast intervals, depending on the specific strategy. The width of these bands is determined by the quantiles of the relevant distributions at a common confidence level $\alpha$, which we refer to as the ``bandwidth level''. 
The bandwidth level $\alpha$ controls the aggressiveness of the strategies by regulating the frequency of trading signals. Larger values of $\alpha$ yield narrower bands and, therefore, more frequent trading. This raises the gross profit potential by exploiting a larger number of spread deviations, but also amplifies cumulative transaction costs and the exposure to short-lived noisy signals. On the contrary, smaller values of $\alpha$ yield wider bands and more conservative trading strategies, reducing transaction costs at the expense of potentially foregone trading opportunities.}

\textcolor{black}{In Subsection \ref{subsec:TestingTheStrategies}, we examine the sensitivity of strategy performance to different choices of $\alpha$, showing that some strategies exhibit robust performance across a wide range of values. For all further experiments, we consider $\alpha = 0.20$, which represents an intermediate choice within the range commonly adopted in the literature, which spans from $\alpha = 0.30$ (bands of approximately one standard deviation) to $\alpha = 0.05$ (bands of approximately two standard deviations).}\\

\textbf{Strategy 1: plain vanilla (PV)}. This first strategy is a benchmark in the pairs trading literature (see \cite{Burgess}). According to this rule, the investor should open a position as soon as the spread differs from zero. This strategy implies that the investor exploits every deviation of the assets from their long-run relationship and might be highly profitable. However, this rule involves very frequent trading (since the spread oscillates frequently around zero) and, as shown below, might not be profitable as soon as transaction costs are taken into account.

Formally, assuming there is no open position at $t-1$, a position is opened at $t$ if $S_t \neq 0$.\\

\textbf{Strategy 2: probability interval (ProbI)}. If the investor prefers avoiding too much trading in order to save on transaction costs, she should open a position only when the spread deviates significantly from zero. One common choice to quantify this deviation is to trade when the spread exceeds the so-called Bollinger band (see, e.g., \cite{ES19}), which coincides with the 95\% probability interval under a normality assumption, where the two moments of the distribution are estimated dynamically over a rolling window of $n$ days. This improves on the related strategy analyzed in the seminal work of \cite{GGR06}, where the bands are just assumed to be constant and equal to twice the sample standard deviation. The present ProbI strategy generalizes the strategy analyzed in \cite{BM09}, who consider two  different bands associated to two alternating market regimes.

Formally, assuming there is no open position at $t-1$, a position is opened at time $t$ if $S_t \notin ( \hat{\mu}_{t-1-n:t-1} \pm | q_{\frac{\alpha}{2}} | \hat{\sigma}_{t-1-n:t-1} )$, where $\hat{\mu}_{n_1:n_2}$ (resp. $\hat{\sigma}_{n_1:n_2}$) is the sample mean (resp. standard deviation) estimated using the datapoints from $n_1$ to $n_2$, $q_{\frac{\alpha}{2}}$ is the $\frac{\alpha}{2}$-quantile of the standard normal distribution and $\alpha$ is the previously described bandwidth level. This strategy is equivalent to the one based on the $z$-score (considered for instance in \cite{AL10}), defined as $z_t := \frac{S_t - \hat{\mu}_{t-1-n:t-1}}{\hat{\sigma}_{t-1-n:t-1}}$, that prescribes to open a position at $t$ if $z_t \notin (-q_{\frac{\alpha}{2}},+q_{\frac{\alpha}{2}})$.\\

\textbf{Strategy 3: prediction interval (PredI)}. 
By exploiting our stochastic model of Section \ref{sec:Cointegration}, we can improve on the previous strategy by adopting a forward-looking approach, inspired by \cite{EHM05}. According to the PredI rule, the interval the investor should consider is not the backward-looking probability interval introduced in strategy ProbI, but rather the prediction interval where the moments are computed according to the stochastic model of Section \ref{subsec:OUHMM}.

Formally, assuming there is no open position at $t-1$, a position is opened at time $t$ if $S_t \notin ( \mathbb{E}[S_t | \mathcal{F}^S_{t-1}] \pm |q_{\frac{\alpha}{2}}| \sqrt{\mathbb{V}ar[S_t | \mathcal{F}^S_{t-1}]} )$, where $\mathbb{E}[S_t | \mathcal{F}^S_{t-1}]$ is the one step-ahead forecast of $S_t$ and $\mathbb{V}ar[S_t | \mathcal{F}^S_{t-1}]$ its variance, both computed by relying on the filtering algorithm described in Section \ref{subsec:filter_EM} on the basis of the information generated by spread process $S$ itself.
More specifically, 
\[
\mathbb{E}[S_t | \mathcal{F}^S_{t-1}] = \hat{\gamma}^{(t-1)}(\hat{\mathbf{X}}_{t-1}) + \hat{\alpha}^{(t-1)}(\hat{\mathbf{X}}_{t-1})S_{t-1}
\qquad\text{and}\qquad
\mathbb{V}ar[S_t | \mathcal{F}^S_{t-1}] = ( \hat{\eta}^{(t-1)}(\hat{\mathbf{X}}_{t-1}) )^2.
\]
Observe that these quantities depend on the filtered estimate of the latent Markov chain $\textbf{X}$.\\

\textbf{Strategy 4: realized increment (RI)}. As suggested by \cite{DLE06}, another way to avoid too much trading and save on transaction costs is to look for increments of the spread that are significantly ``larger than usual''. According to this strategy, the investor has to first compute the time series of the spread increments, that we denote by $x=(x_t)_{t \in \mathbb{N}}$ with $x_t := S_t/S_{t-1} -1$. Then, she should open a position at time $t$ if the increment $x_t$ is significantly larger than the previous ones. A standard way to formalize this is to rely on two empirical quantiles. 

Formally, assuming there is no open position at $t-1$, a position is opened at time $t$ if $x_t \notin ( - |q_{\frac{\alpha}{2},x} |, |q_{\frac{\alpha}{2},x}| )$ where $q_{\frac{\alpha}{2},x}$ is the empirical quantile of order $\frac{\alpha}{2}$ of the spread increment $x$.\\

\textbf{Strategy 5: predicted increment (PI)}. 
By exploiting the stochastic model of Section \ref{sec:Cointegration},  we can improve on strategy RI described above. More specifically, the investor can compute the series $\hat{x}=(\hat{x}_t)_{t \in \mathbb{N}}$ of predicted spread increments, with $\hat{x}_t := \mathbb{E}[S_{t+1} | \mathcal{F}^S_{t}]/S_t -1$, and trade whenever the predicted increment lies outside a given interval.

Formally, assuming there is no open position at $t-1$, a position is opened at time $t$ if $\hat{x}_t \notin ( - |q_{\frac{\alpha}{2},x} |, |q_{\frac{\alpha}{2},x}| )$.\\

Finally, we shall also consider two passive strategies that do not incur into transaction costs: a buy-and-hold position on the S\&P500 index (Ex. S\&P), which will be used as a generic benchmark, and a buy-and-hold position on the Invesco DB Oil Fund\footnote{See \hyperlink{https://www.invesco.com/us/financial-products/etfs/product-detail?audienceType=Investor&ticker=DBO}{https://www.invesco.com/us/financial-products/etfs/product-detail?audienceType=Investor\&ticker=DBO}. Numerical experiments show that different alternatives, like the United States Oil Fund ETF 
or the WisomTree Crude Oil ETF, 
are characterized by almost identical performances.} (Ex. ETF), which is one of the most liquid ETFs for crude oil futures markets and represents a simple alternative for an investor willing to invest in the crude oil futures market. For sake of comparison, the returns of both these strategies are computed in excess of the prevailing risk-free rate.

\begin{remark}[Alternative statistical arbitrage strategies]
    Other statistical arbitrage strategies inspired by pairs trading proposed in the literature involve: moving-average trading strategies based on the difference between a short and a long moving average for the spread (see \cite{AN08}); strategies based on first-passage times when the spread is modelled as a standard Ornstein-Uhlenbeck process (see \cite{Ber10} and \cite{CB12}); strategies constructed by relying on a machine learning approach that compares possible investment strategies on a training sample (see \cite{SarmentoHorta}).
\end{remark}

\subsection{Performance measurement}\label{subsec:PerformanceMeasure}

As pointed out in \cite{GGR06}, computing returns for trading strategies based on the concept of pairs trading is non-trivial due to several factors that complicate the calculation and interpretation of performance metrics. First of all, there are days in which a trading position is open and days in which there is no open position. Moreover, a successfully executed trade delivers a positive cashflow when opened and a non-negative cashflow when closed, thus preventing the computation of a standard return. 
To overcome these issues, it is common to rely on the daily mark-to-market profit \& loss indicator.

Let us assume that the strategy is applied over a time period spanning $n$ days. Following the original methodology proposed by \cite{GGR06}, which is widely used in the context of pairs trading strategies (see, e.g., \cite{CB12} and \cite{HainHessUhrigHomburg2018}), we compute the daily mark-to-market return $r_t$ for each day $t = 1, \ldots, n$ as
\begin{equation}\label{eqn:Definitionr}
    r_t = \sum_{k=1}^K \left( I_t^k \cdot \frac{F_t^k - F_{t-1}^k}{F_{t-1}^k} \cdot \frac{F_{t-1}^k | \lambda^k |}{| \lambda^0| +\sum_{j=1}^K F_{t-1}^j | \lambda^j |}
    - | \Delta I_{t}^k | \cdot \frac{c_k F_t^k | \lambda^k |}{| \lambda^0| +\sum_{j=1}^K F_{t-1}^j | \lambda^j |} \right)
\end{equation}
where:
\begin{itemize}
    \item $K$ is the number of futures contracts considered in the trading strategy, which is three in our benchmark case;
    \item $I_t^k$ is a three-state indicator variable that takes the value $1$, $-1$, or $0$, corresponding to a long, short, or neutral (no position) state, respectively, for futures contract $k$. $\Delta I_t^k = I_t^k - I_{t-1}^k$, which is nonzero only when a position is opened or closed on day $t$;
    \item $c_k$ represents the transaction cost parameter associated with trading futures contract $k$, applied when a position is opened or closed.
\end{itemize}
\textcolor{black}{The first term in \eqref{eqn:DefinitionR} represents the daily mark-to-market profit or loss generated by the futures positions, weighted by their relative contribution to the total gross exposure. The second term accounts for transaction costs incurred when positions are opened or closed, proportional to the current market value of the traded contracts and independent of trading direction.
By construction, if there is no open position on day $t$, then $r_t = 0$.
The daily returns defined in \eqref{eqn:DefinitionR} represent excess returns per unit of gross notional exposure,  as in \cite{HainHessUhrigHomburg2018}. 
In particular, the normalization ensures comparability across different strategies.
Because trading positions are initiated only when the spread deviates sufficiently from its equilibrium value, the opening of a position generates an immediate positive cash inflow, while closing positions produces no negative cash outflow (apart from transaction costs). As a consequence, the daily returns defined in \eqref{eqn:DefinitionR} can be interpreted as excess returns.}

The overall performance of a given strategy is then evaluated by computing 
\begin{equation}\label{eqn:DefinitionR}
    R := \prod_{t = 1}^n (1 + r_t) -1.
\end{equation}
To facilitate comparison across samples of different lengths, we shall always report the returns in (\ref{eqn:DefinitionR}) in annual terms: following the usual convention, this amounts to multiply $R$ by $250/n$.

\textcolor{black}{Daily returns computed as above typically alternate between periods in which a position is active and periods in which no position is held. As a consequence, daily returns may exhibit substantial short-term variability and are of limited standalone relevance for an investor primarily concerned with the profitability over complete trading episodes. For this reason, we introduce an additional trade-level performance measure based on individual trading windows, defined as the time interval between the opening and closing of a position.
Formally, let $I^k_t \in \{-1,0,1\}$ denote the position indicator for futures contract $k$ at time $t$, as defined above, and define the aggregate position indicator
\[
I_t := \max_{k=1,\dots,K} |I^k_t|,
\]
which equals one whenever at least one futures position is open and zero otherwise. 
Let $\{(\sigma_j,\tau_j)\}_{j=1}^{N_{\mathrm{TW}}}$ denote the resulting sequence of non-overlapping trading windows, where $N_{\mathrm{TW}}$ denotes the total number of completed trading windows over the entire sample and
\[
\sigma_j := \min\{t>\tau_{j-1} : I_t = 1,\ I_{t-1} = 0\}
\quad \text{ and } \quad
\tau_j := \min\{t > \sigma_j : I_t = 0\},
\]
with $\tau_0:=0$.
For each trade $j=1,\ldots,N_{\mathrm{TW}}$, the trading-window return is defined as
\begin{equation}\label{eqn:RTW}
R^{\mathrm{TW}}_j
:= \prod_{t=\sigma_j+1}^{\tau_j} \bigl(1+r_t\bigr) - 1,
\end{equation}
where $r_t$ denotes the daily return defined in \eqref{eqn:Definitionr}. 
For brevity of notation, we denote by $R^{\mathrm{TW}}$ the collection of trading-window returns over the entire sample. Note that, since $r_t = 0$ whenever no position is open, $R^{\mathrm{TW}}_j$ coincides with the compounded return generated over the holding period of trade $j$. Trading-window returns thus provide a direct measure of per-trade profitability, complementary to daily and aggregate return measures.}

When evaluating the performance of statistical arbitrage strategies, which typically entail frequent trading, transaction costs can play a significant role and substantially affect net profitability, as discussed in \cite{DF12}. Following \cite{HainHessUhrigHomburg2018}, we estimate each contract-specific transaction cost parameter $c_i$ as the average effective daily half bid-ask spread over the full sample period. The resulting transaction cost estimates are reported in Table \ref{tab:transactionCosts}. As expected, the highest transaction cost is associated with the Shanghai contract, which corresponds to the youngest and least mature market among those considered and may therefore be subject to relatively higher trading frictions.
\begin{table}[]
        \centering
        \begin{tabular}{cc}
             Futures contract & Transaction costs parameter $c_i$ (bps) \\ \hline
             Brent & 5.80 \\
             Shanghai & 53.71 \\
             WTI & 20.24 \\
             Dubai & 1.52
        \end{tabular}
        \caption{Estimates of the transaction cost parameters $c_i$ for the four futures contracts considered in our statistical arbitrage strategies.}
        \label{tab:transactionCosts}
    \end{table}
Overall, the estimated transaction cost parameters are in line with values documented in the literature. In particular, studies on pairs trading involving USD-denominated crude oil futures report transaction costs ranging from approximately $c = 20$ bps in \cite{AN08} to about $c = 60$ bps in \cite{DF12}.

Unlike pure arbitrage opportunities, statistical arbitrage strategies are not exempt of financial risk. For this reason, we  also consider the Sharpe ratio, a widely used risk-adjusted performance measure
\begin{equation}\label{eqn:DefinitionSR}
    SR := \frac{\frac{1}{n} \sum_{t=1}^n r_t}{\sqrt{\frac{1}{n} \sum_{t=1}^n \left(r_t - \frac{1}{n} \sum_{v=1}^n r_v \right)^2 }},
\end{equation}
where we do not subtract the prevailing risk-free rate in the numerator, as our returns $r_t$ represent excess returns. For the sake of comparability, we shall always express the Sharpe Ratio of a given trading strategy in annual terms. 
In Section \ref{subsec:VaRAnalysis}, we will perform an additional analysis by also computing the Value-at-Risk, the Expected Shortfall and the Maximum Drawdown of the different strategies introduced in Section \ref{subsec:ParisTradingStrategies}. Moreover, to address the limitation that the Sharpe Ratio captures only the first two moments of the return distribution, we also examine the full empirical distribution of returns generated by our strategies.

\section{Empirical analysis}\label{sec:Empirics}

This section is divided into four subsections. In Section \ref{subsec:AnIllustrativeExample}, we perform the cointegration analysis among the three futures prices over a fixed time period. In Section \ref{subsec:TestingTheStrategies}, we evaluate and compare the performance of the statistical arbitrage strategies introduced in Section \ref{subsec:ParisTradingStrategies}. 
In Section \ref{subsec:DifferentUnderlyings}, we test our strategies with respect to different choices of the futures contracts, replacing the Shanghai futures with the Dubai futures.
Finally, in Section \ref{subsec:validationOfTheResults}, we verify the robustness of our findings by testing the statistical arbitrage strategies with respect to different levels of transaction costs and over different time periods. 

\subsection{Cointegration analysis, filtering and parameter estimation}\label{subsec:AnIllustrativeExample}

\subsubsection{Data description}\label{Data description}
We consider daily\footnote{Statistical arbitrage opportunities arise also intraday. See for instance \cite{MarshallNguyenVisaltanachoti2013} for an analysis of this kind of opportunities in the ETFs market.} and weekly futures prices for the Brent, the WTI and the Shanghai futures, the latter converted from CNY to USD. Specifically, we focus on continuous price series constructed from futures contracts with the nearest monthly maturity, identified by the following Datastream mnemonics: \texttt{LLCC.01} for Brent, \texttt{NCLC.01} for WTI, \texttt{SICC.01} for Shanghai, and \texttt{OLDUB1M} for Dubai crude oil futures, which is introduced in the analysis in Subsection \ref{subsec:DifferentUnderlyings}. By construction, these rolled series may exhibit a non-negligible roll yield, arising from the fact that the price of an expiring contract can differ from that of the next-nearest maturity. However, as discussed in Section \ref{sec:Cointegration}, our statistical arbitrage strategies are based on the cointegrating relationships among these futures. Since all the futures contracts are written on essentially the same underlying asset (crude oil), the estimated cointegrating coefficients tend to sum to a value close to zero. This property naturally mitigates the impact of roll yield, as the roll yields across different contracts usually move together and, when combined linearly with cointegrating coefficients that approximately sum to zero, their effects largely offset one another. 

\begin{figure}
    \centering
    \includegraphics[width=\textwidth]{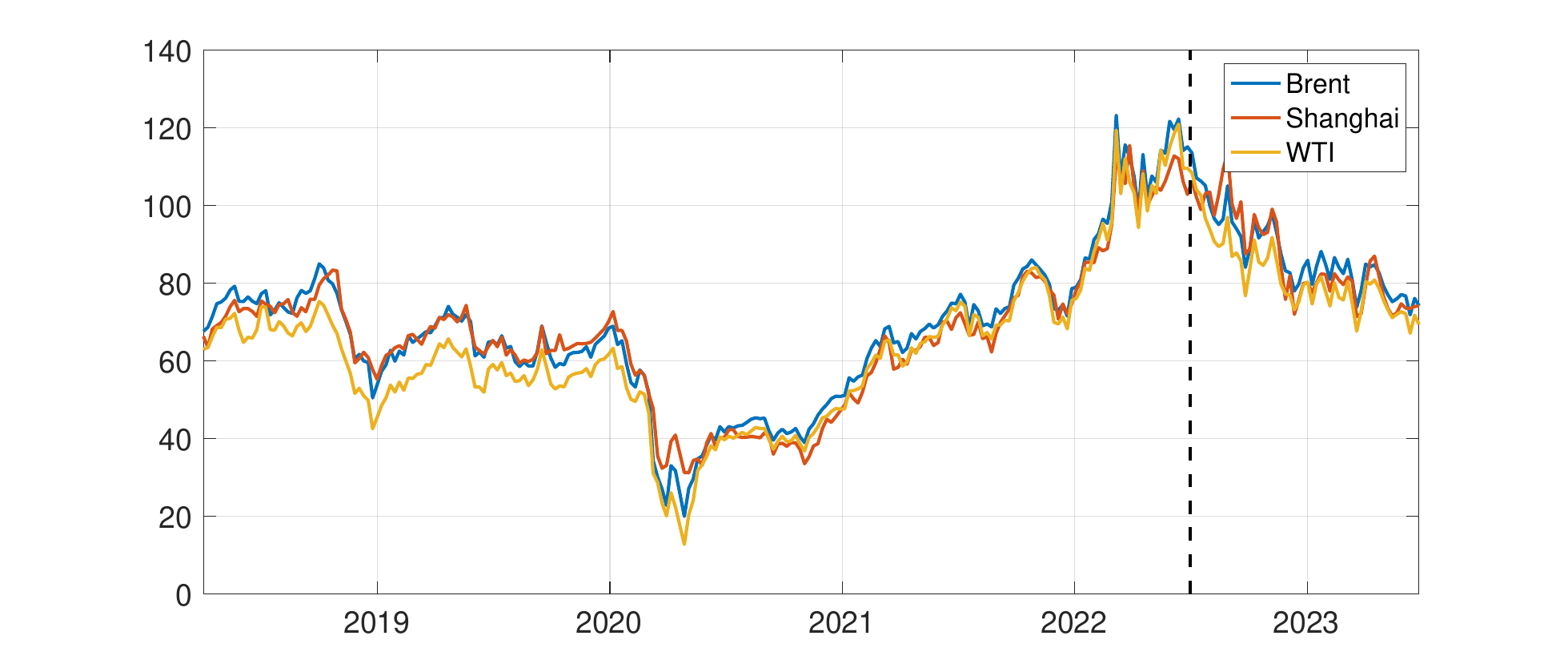}
    \caption{Weekly futures prices of the three contracts. The black dashed line corresponds to 07/01/2022, which separates the training sample from the test sample.}
    \label{fig:futuresPrices_BSW}
\end{figure}

While the time series for the Brent and the WTI start in the eighties, the Shanghai has been trading only since March 2018. Therefore, our dataset spans from $t_0=$ 03/26/2018 to $T=$ 06/30/2023, including 1373 daily observations and 273 weekly ones. The weekly futures prices are shown in Figure \ref{fig:futuresPrices_BSW}, where the sharp decline in crude oil prices following the COVID-19 crisis in early 2020 as well as the pronounced peak associated with the outbreak of the Russia-Ukraine conflict in 2022 and the subsequent period of increased volatility are clearly visible.

We split the full dataset into a \textit{training sample} and a \textit{test sample}. We label by $t_B$ the first date of the test sample, which in our analysis is set at 07/01/2022, represented by a black dashed line in Figure \ref{fig:futuresPrices_BSW}. Therefore, the test sample includes a year of observations, corresponding to 261 (resp. 52) daily (resp. weekly) observations.

\subsubsection{Unit roots, VAR and cointegration analysis}\label{subsec:UnitRootsVARCointegration}

We perform a cointegration analysis based on weekly data, following the usual steps of cointegration analysis (see, e.g., \cite[Chapter~4]{GuidolinPedio}). The choice of using weekly data is typical in the literature as daily data tend to be more volatile and noisy due to several factors such as intraday trading, news releases, and other short-term fluctuations. In the  Supplementary Materials, we report the results of the cointegration analysis on daily data: while our main findings are not affected by the choice of weekly or daily data, it turns out that the performances of our statistical arbitrage strategies are superior when weekly data are used for the cointegration analysis. 

Before assessing the presence of one or more cointegrating relationships among the three futures prices, we test whether each of them is integrated of some order. To this effect, we run standard unit root/stationarity tests on each of the three series individually, running the ADF and the PP tests (\cite{SD84}, \cite{PP88}) for the presence of a unit root. Moreover, we account for possible structural breaks at an unknown time by running also the breakpoint Perron unit root test (\cite{P97}). The $p$-values of these tests are reported in Table \ref{tab:ADF_PP_tests_BSW}.
\begin{table}[]
    \centering
    \begin{tabular}{c|ccc}
            & Brent & Shanghai & WTI\\ \hline
        ADF & 0.9434 & 0.8535 & 0.9423\\
        PP & 0.8788 & 0.8413 & 0.8714\\
        bP & 0.9548 & 0.9471 & 0.9425
    \end{tabular}
    \caption{$p$-values of the Augmented Dickey-Fuller (ADF), the Phillips-Perron (PP) and the breakpoint Perron (bP) tests. The number of lag length for the ADF and the bP is chosen according to the Schwarz IC.}
    \label{tab:ADF_PP_tests_BSW}
\end{table}
We find a strong evidence that the three series contain (at least) a unit root. The same tests on the first differences of the three series deliver $p$-values all lower or equal to 1\%, indicating that the series contain only one unit root. Therefore, we can conclude that the three series are integrated of order one.

Given the similarities among the three crude oil futures contracts, it is natural to study whether the stochastic trend found in the three series is common. Among the two main approaches to cointegration, the one of \cite{EG87} and the one of \cite{J88} and \cite{Johansen}, we opt for the latter. Indeed, the Engle and Granger approach would require to choose a priori a reference futures contract among the three under consideration and there is no objective way to do so. In order to assess any cointegrating relationship by means of the Johansen procedure, we first formulate a vector autoregressive model of order $p$ (VAR($p$) henceforth). We choose $p$ by minimizing the Schwarz IC, which is known to deliver the most parsimonious specification. On our test sample, we obtain that the optimal number of lags $p$ is equal to two (this result is quite stable even when considering different test samples, namely when changing $t_0$ and/or $t_B$). 
The resulting (reduced-form) VAR(2) can be written as
$$
\mathbf{F}_t = \mathbf{a}_0 + \mathbf{A}_1 \mathbf{F}_{t-1} + \mathbf{A}_2 \mathbf{F}_{t-2} + \mathbf{u}_t
$$
where $\mathbf{F}_t = (F_t^B, F_t^S, F_t^W)$ are the three variables of interest, $\mathbf{a}_0 \in \mathbb{R}^3$, $\mathbf{A}_i \in \mathbb{R}^{3 \times 3}$, $i=1,2$, are the parameters of the model and $\mathbf{u}_t$ is a three-dimensional white noise process with covariance matrix $\mathbf{\Sigma} \in \mathbb{R}^{3 \times 3}$. The VAR(2) model is estimated by ordinary least squares and the resulting point estimates and standard errors are given by
\begin{align*}
    \left[ \begin{array}{c}
F_t^B \\ F_t^S \\ F_t^W
\end{array} \right] & = \left[ \begin{array}{c}
\underset{(1.095)}{1.546} \\ \underset{(0.883)}{1.111} \\ \underset{(1.093)}{1.354}
\end{array} \right] + \left[ \begin{array}{ccc}
\underset{(0.249)}{0.614^{**}} & \underset{(0.125)}{0.088} & \underset{(0.234)}{0.147}
\\ \underset{(0.201)}{0.510^{**}} & \underset{(0.101)}{0.555^{***}} & \underset{(0.189)}{-0.100} \\
\underset{(0.249)}{-0.221} & \underset{(0.125)}{0.111} & \underset{(0.233)}{0.968^{***}}
\end{array} \right] \left[ \begin{array}{c}
F_{t-1}^B \\ F_{t-1}^S \\ F_{t-1}^W
\end{array} \right] \\
&\quad + \left[ \begin{array}{ccc}
\underset{(0.258)}{0.266} & \underset{(0.113)}{-0.157} & \underset{(0.238)}{0.034} \\
\underset{(0.208)}{-0.221} & \underset{(0.091)}{0.158^{*}} & \underset{(0.192)}{0.077} \\
\underset{(0.257)}{0.242} & \underset{(0.112)}{-0.214^{*}} & \underset{(0.237)}{0.100}
\end{array} \right] \left[ \begin{array}{c}
F_{t-2}^B \\ F_{t-2}^S \\ F_{t-2}^W
\end{array} \right] + \left[ \begin{array}{c}
u_{t}^B \\ u_{t}^S \\ u_{t}^W
\end{array} \right]
\end{align*}
where $^{***}$, $^{**}$ and $^{*}$ represent, respectively, a statistical significance at the 1\%, 5\% and 10\% level. Inspecting the roots of the characteristic equation, this VAR(2) specification appears to satisfy the stability condition.

We can now perform the cointegration test. We adopt the trace test of \cite{J88} and allow for a constant in the cointegrating relationship.
\begin{table}[]
    \centering
    \begin{tabular}{ccccc}
         r & h & Trace test stat. & Crit. val. & $p$-value \\ \hline
         0 & 1 & 35.5106 & 29.7976 & 0.0099 \\
         1 & 0 & 10.0981 & 15.4948 & 0.3074 \\
         2 & 0 & 0.8131 & 3.8415 & 0.5153 \\
    \end{tabular}
    \caption{Results of \cite{J88} trace test. $h$ is the rejection decision of the trace test with null hypothesis ``there exist less than or equal to $r$ cointegrating relationships''.}
    \label{tab:CointegrationTests}
\end{table}
Looking at the results reported in Table \ref{tab:CointegrationTests}, the test indicates the presence of only one cointegrating relationship. Moreover, in our dataset we find that this result is robust also when considering different time windows. We can therefore estimate the following vector error correction model (VECM henceforth):
$$
\Delta \mathbf{F}_t = \pmb{\pi}_0 + \mathbf{\Pi}_0 \mathbf{F}_{t-1} + \mathbf{\Pi}_1 \Delta \mathbf{F}_{t-1} + \mathbf{\Pi}_2 \Delta \mathbf{F}_{t-2} + \pmb{\varepsilon}_t
$$
where $\Delta$ is the first difference operator and $\pmb{\pi}_0 \in \mathbb{R}^3$, $\mathbf{\Pi}_i \in \mathbb{R}^{3 \times 3}$, $i=0,1,2$, $c \in \mathbb{R}$ are the parameters of the model and $\pmb{\varepsilon}_t$ is a three-dimensional white noise process with covariance matrix $\mathbf{\Sigma}_u \in \mathbb{R}^{3 \times 3}$.

Since we cannot reject that $\mathrm{rank} \left[ \mathbf{\Pi}_0 \right] = 1$, the matrix $\mathbf{\Pi}_0$ can be decomposed as $\mathbf{\Pi}_0 = \pmb{\alpha} \pmb{\beta}^T$ with $\pmb{\beta} \in \mathbb{R}^3$. Rewriting $\pmb{\pi}_0$ as $\pmb{\pi}_0 = \pmb{\alpha} c_0 + \mathbf{c}_1$, with $c_0 \in \mathbb{R}$ and $\mathbf{c}_1 \in \mathbb{R}^3$, we can derive the expression of the resulting cointegration relationship, $c_0 + \pmb{\beta}^\top{} \mathbf{F}_t$, which is our spread process $S$, and the adjustment coefficients $\pmb{\alpha}$ that measure the speed of convergence of each series to the long-run relationship. \textcolor{black}{The maximum likelihood estimates of the full VECM, together with their standard errors and statistical significance, indicated as before by $^{***}$, $^{**}$, and $^{*}$, are}
\textcolor{black}{\begin{align}
\left[
\begin{array}{c}
\Delta F_t^B \\ \Delta F_t^S \\ \Delta F_t^W
\end{array}
\right]
&=
\left[
\begin{array}{c}
\underset{(0.271)}{0.200} \\
\underset{(0.217)}{0.131} \\
\underset{(0.269)}{0.199}
\end{array}
\right]
+
\left[
\begin{array}{c}
\underset{(0.151)}{0.059} \\
\underset{(0.121)}{0.419^{***}} \\
\underset{(0.149)}{0.063}
\end{array}
\right]
\Big(
F_{t-1}^B
- 0.698\,F_{t-1}^S
- 0.340\,F_{t-1}^W
+ 0.432
\Big) \nonumber
\\
&\quad
+
\left[
\begin{array}{ccc}
\underset{(0.267)}{-0.362} & \underset{(0.142)}{0.088} & \underset{(0.235)}{0.140} \\
\underset{(0.214)}{0.109}  & \underset{(0.114)}{-0.162} & \underset{(0.188)}{0.039} \\
\underset{(0.265)}{-0.220} & \underset{(0.141)}{0.093} & \underset{(0.234)}{0.007}
\end{array}
\right]
\left[
\begin{array}{c}
\Delta F_{t-1}^B \\ \Delta F_{t-1}^S \\ \Delta F_{t-1}^W
\end{array}
\right] \label{eqn:fullVECM}
\\
&\quad
+
\left[
\begin{array}{ccc}
\underset{(0.263)}{-0.132} & \underset{(0.119)}{-0.073} & \underset{(0.239)}{0.267} \\
\underset{(0.211)}{-0.103} & \underset{(0.095)}{-0.133} & \underset{(0.191)}{0.144} \\
\underset{(0.261)}{0.111}  & \underset{(0.118)}{-0.102} & \underset{(0.237)}{0.061}
\end{array}
\right]
\left[
\begin{array}{c}
\Delta F_{t-2}^B \\ \Delta F_{t-2}^S \\ \Delta F_{t-2}^W
\end{array}
\right]
+
\left[
\begin{array}{c}
\varepsilon_t^B \\ \varepsilon_t^S \\ \varepsilon_t^W
\end{array}
\right]. \nonumber
\end{align}}
\textcolor{black}{Having framed the system within a VECM framework, it is possible to investigate both short-run and long-run Granger causality, as well as their direction, among the variables. The estimates reported in \eqref{eqn:fullVECM} provide limited evidence of short-run causality, since the coefficients on the lagged first-difference terms are generally not statistically significant at conventional levels. In contrast, the adjustment coefficient in the equation for $\Delta F_t^{S}$ is positive and statistically significant at the 1\% level, while the corresponding coefficients in the equations for $\Delta F_t^{B}$ and $\Delta F_t^{W}$ are not statistically different from zero. This pattern indicates that deviations from the long-run equilibrium are primarily corrected through adjustments in $F_t^{S}$, implying a unidirectional long-run causal relationship running from $F_t^{B}$ and $F_t^{W}$ to $F_t^{S}$. From a trading perspective, the fact that the Shanghai futures price adjusts to restore the long-run relationship suggests that it plays a stabilizing role within the system, making it a promising candidate for pairs trading strategies based on mean reversion.}

\begin{figure}
    \centering
    \includegraphics[width=\textwidth]{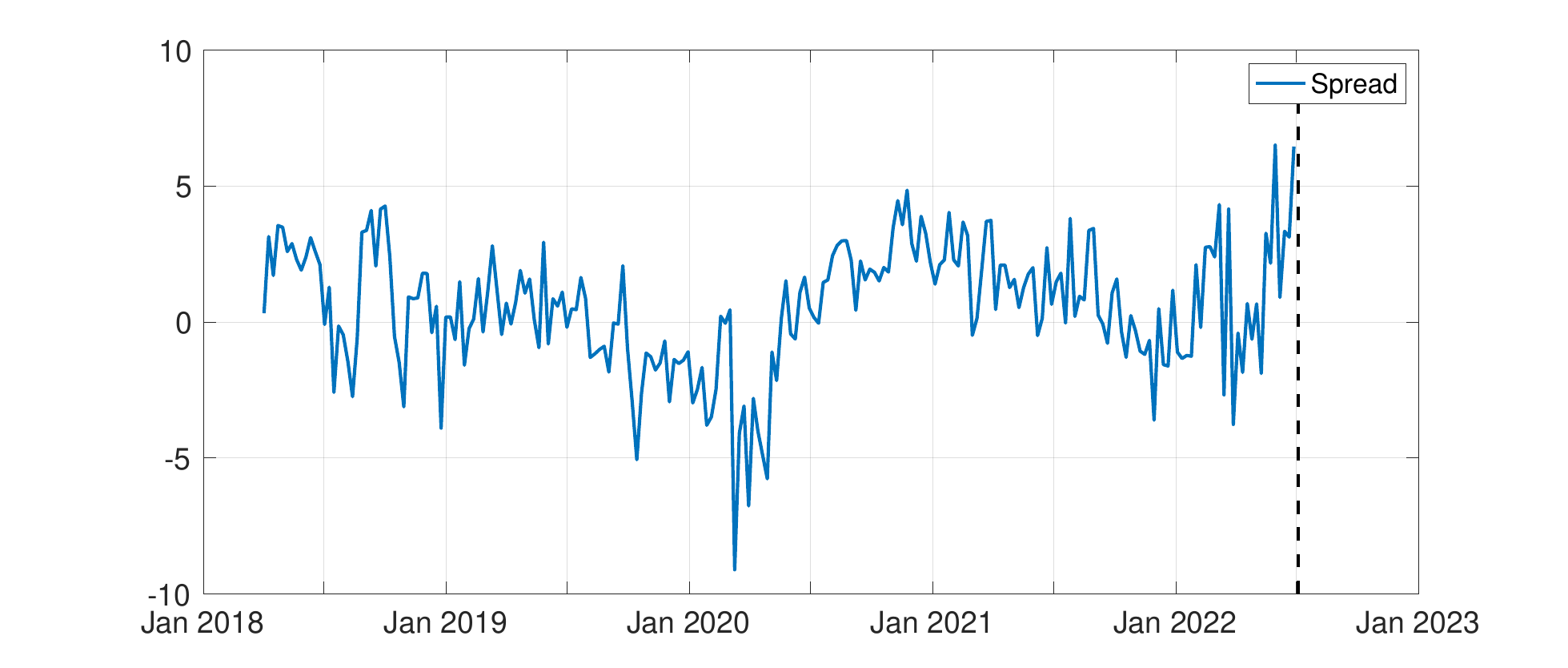}
    \caption{Resulting spread process within the test sample.}
    \label{fig:WeeklySpread}
\end{figure}

\textcolor{black}{Given the estimates in (\ref{eqn:fullVECM})}, in our test sample the spread process results to be given by
\begin{equation}\label{eqn:weeklySpreadWeights}
    S_t = F_t^B -0.6982 F_t^S -0.3402 F_t^W +0.4322
\end{equation}
and it is plotted in Figure \ref{fig:WeeklySpread}.
As can be seen from this figure, the spread process exhibits a stationary behavior. Coherently, the ADF and PP tests reject the hull hypothesis of the presence of a unit root while the KPSS test (\cite{KPSS92}) cannot reject the null hypothesis of stationarity of the series.
\begin{figure}
    \centering
    \includegraphics[width=0.48\textwidth]{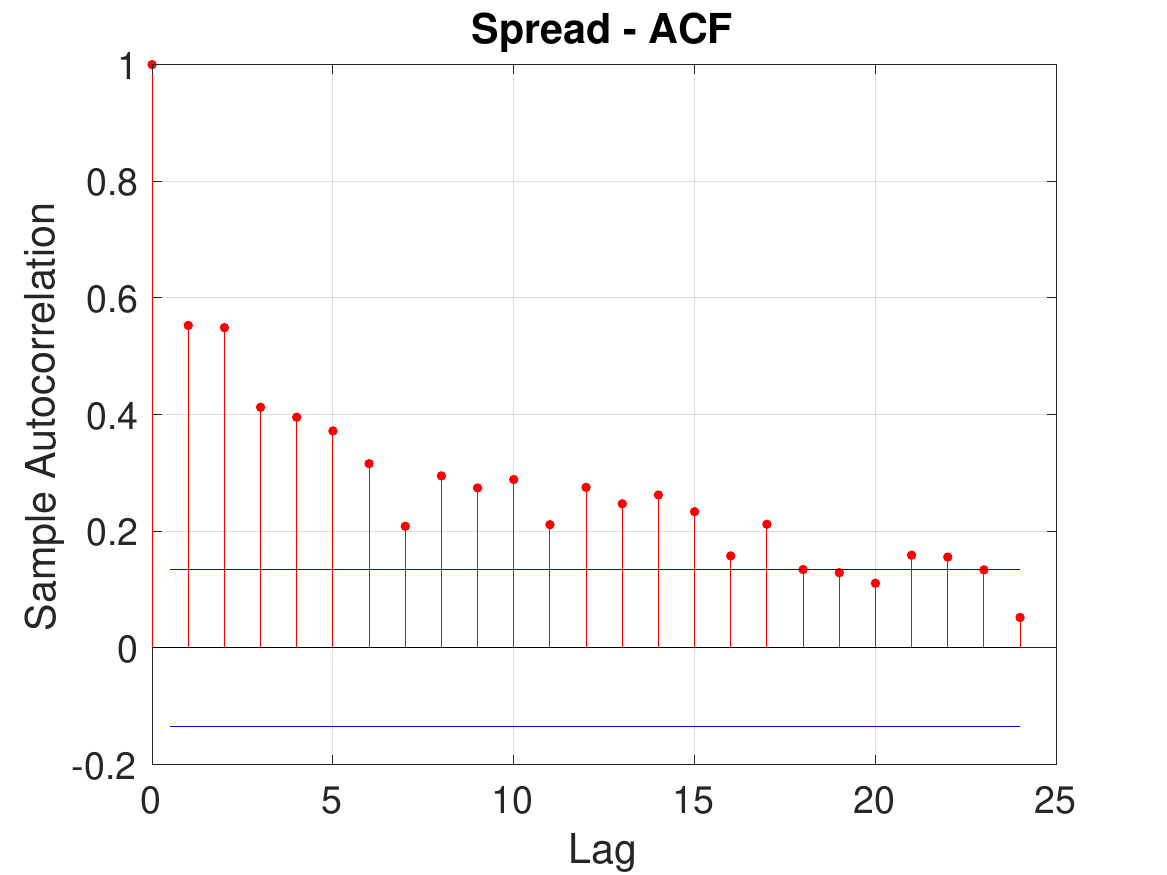}
    \includegraphics[width=0.48\textwidth]{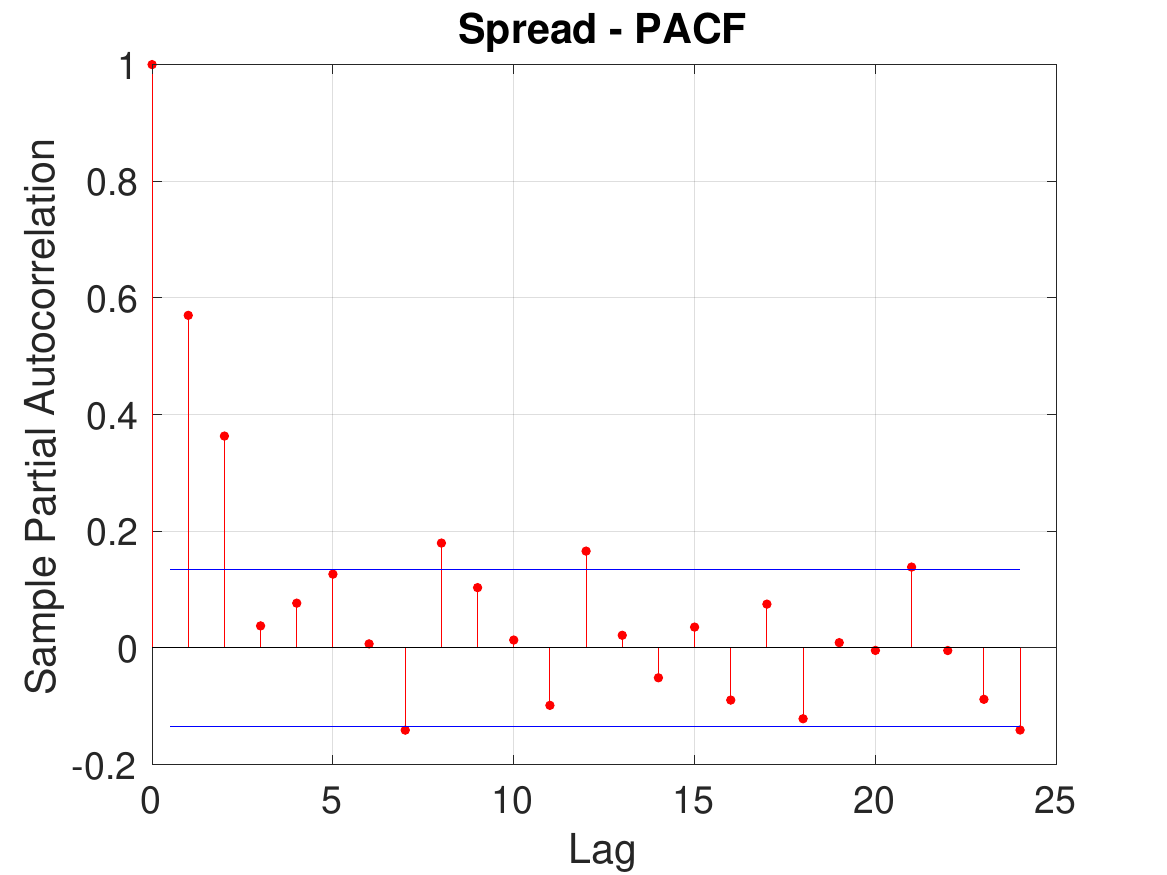}
    \caption{ACF and PACF of the spread process over the training sample, weekly observations.}
    \label{fig:ACF_PACF_weeklySpread}
\end{figure}
Moreover, as can be seen from Figure \ref{fig:ACF_PACF_weeklySpread}, both the ACF and the PACF of $S$ decline to zero. In particular, only the first two lags of the PACF seem to be statistically different from zero. This indicates some degree of mean-reversion of $S$ that empirically justifies modelling $S$ by means of the stochastic model introduced in Section \ref{subsec:OUHMM}.

\subsubsection{Filtering the spread}\label{subsec:FilteringTheSpread}
We now estimate the parameters of the AR-HMM for $S$. We first determine the most likely number $N$ of the states of the latent Markov chain $\textbf{X}$; we then compute daily filtered estimates of $\textbf{X}$; finally, we estimate the model parameters $\pmb{\gamma}$, $\pmb{\alpha}$, $\pmb{\eta}$ in (\ref{eqn:discreteTimeSpread}) and the transition matrix $\pmb{\Pi}$ by means of the filter-based EM algorithm described in Section \ref{subsec:filter_EM}.
We implement the filtering and parameter estimation procedure on daily data. The choice of daily data is motivatied by two reasons: first, we want to rely on a large dataset for the estimation of the parameters of the spread process, in order to capture its short-term variations; second, we want to possibly implement trading strategies on a daily basis rather than on a weekly basis.

The daily spread process is determined by using the weights reported in (\ref{eqn:weeklySpreadWeights}). We estimate the model for $N \in \{1,2,3\}$. Since our trading strategy is based on the forecasts generated by the model, we compute the mean forecast error over the trading sample and select the value of $N$ that minimizes it. According to this criterion, the preferred specification in the training sample features $N=2$ states for the Markov chain $\mathbf{X}$. 

\begin{figure}
    \centering
    \includegraphics[width=\textwidth]{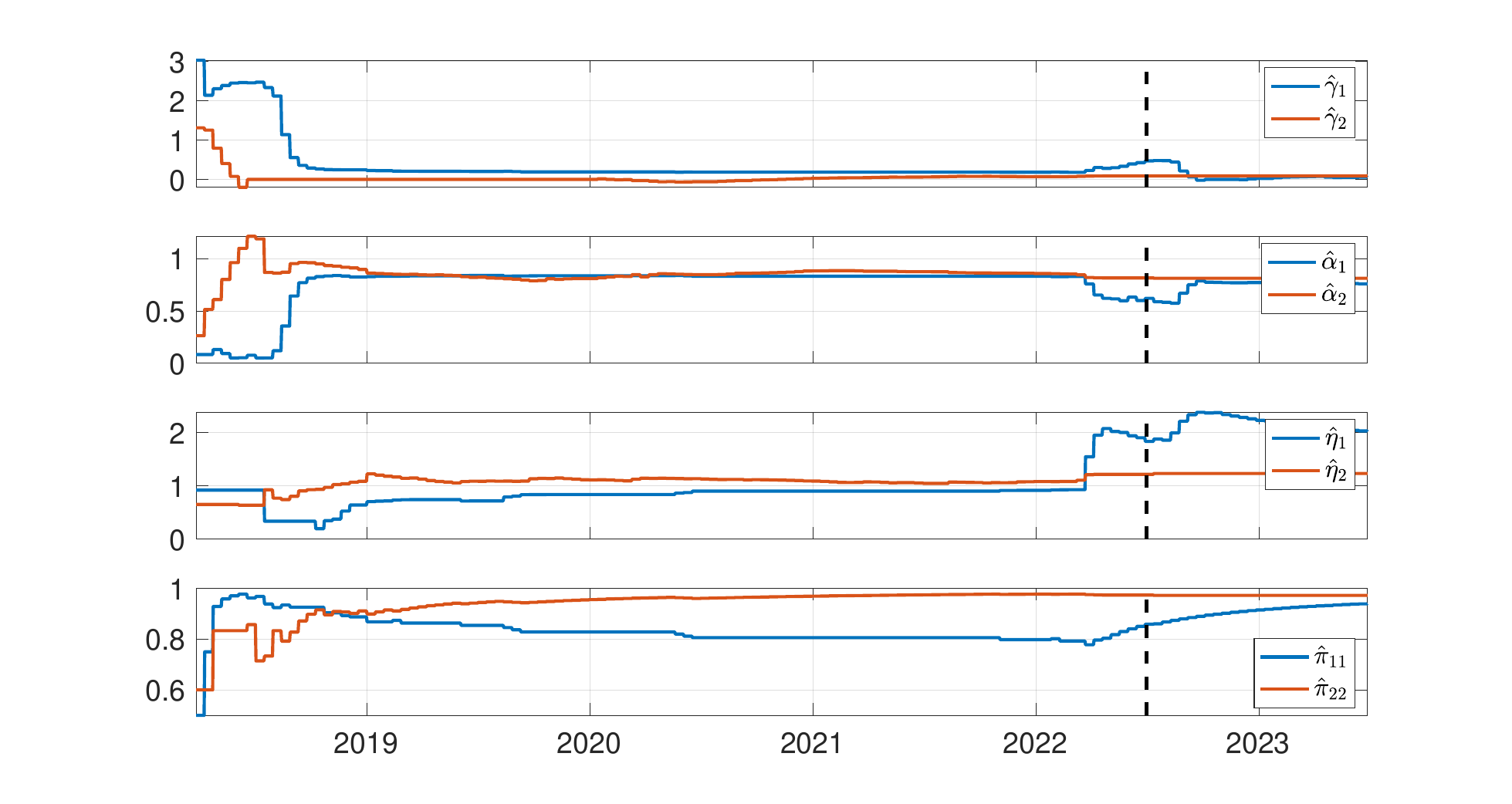}
    \caption{Recursively updated estimates of $\gamma_i$, $\alpha_i$, $\eta_i$ and $\pi_{ii}$, $i=1,2$, for the $N=2$ AR-HMM.}
    \label{fig:DailySpreadFunctionEstimates}
\end{figure}

Figure \ref{fig:DailySpreadFunctionEstimates} illustrates the time evolution of the filtered estimates of the model parameters $\pmb{\gamma}$, $\pmb{\alpha}$, $\pmb{\eta}$ and of the probabilities $\pi_{ii}$, $i=1,2$, of the Markov chain $\textbf{X}$ to remain in its current state.
Due to the well-known issue of slow convergence of the EM algorithm, the estimates turn out to be unstable at the beginning of the training sample. Afterwards, they stabilize and only major changes in the underlying observations have an impact on the estimates. For example, we observe an instability in the filtered parameter estimates around the onset of the war in Ukraine, coinciding with a period of heightened market volatility.

For the sake of reference, row ``baseline'' of Table \ref{tab:estimatesFiltering} reports the parameter estimates at the end of the training sample. The results indicate that both regimes are highly persistent, with the second regime being particularly so, as reflected by a transition probability $\pi_{22}$ close to one. The two regimes exhibit comparable mean-reversion speeds, as measured by the autoregressive coefficients $\alpha$, while differing markedly in their long-run mean and volatility parameters. In particular, regime 1 is characterized by higher values of the long-run mean level $\gamma$ and volatility $\eta$, whereas regime 2 exhibits markedly lower values for both parameters.

\begin{figure}
    \centering
    \includegraphics[width=\textwidth]{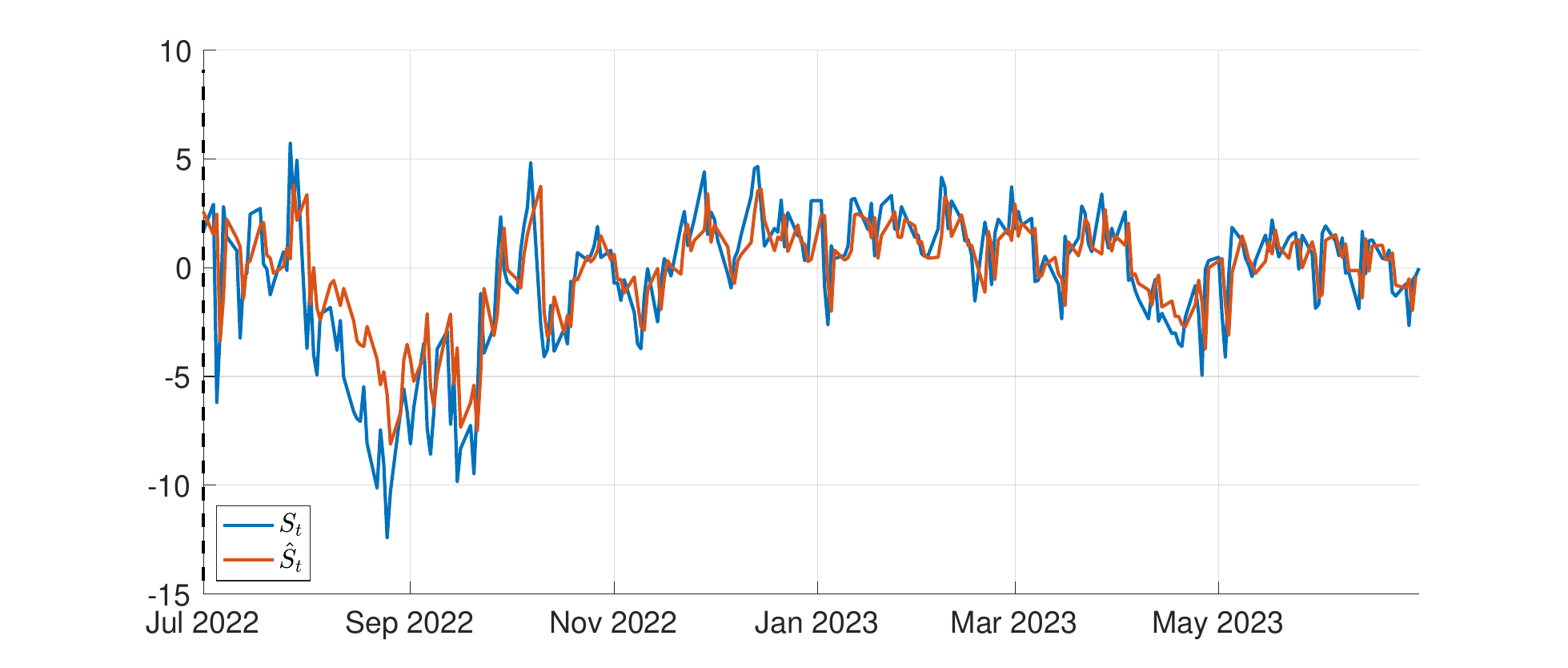}
    \caption{One-step ahead forecasts and actual values of $S$ across the test sample.}
    \label{fig:FilteredValues}
\end{figure}

Figure \ref{fig:FilteredValues} displays the one-step ahead forecast of $S$ along with the realized value across the test sample.
As can be seen from this figure, the filtering algorithm produces fairly good forecasts of the realizations of the spread.

\begin{table}[]
    \centering
    \textcolor{black}{
    \begin{tabular}{c|cc|ccc|ccc}
    \multicolumn{3}{c}{}&\multicolumn{3}{c}{State 1}&\multicolumn{3}{c}{State 2}\\
&$\hat{\pi}_{11}$&$\hat{\pi}_{22}$&$\hat{\gamma}_1$&$\hat{\alpha}_1$&$\hat{\eta}_1$&$\hat{\gamma}_2$&$\hat{\alpha}_2$&$\hat{\eta}_2$ \\ \hline 
baseline&0.8588&0.9732&0.4705&0.6214&1.8343&0.0855&0.8164&1.2187\\
mean&0.9255&0.9627&0.4504&0.6426&1.5606&0.0600&0.8049&1.2777\\
std. dev.&0.0583&0.0559&0.0948&0.1319&0.4821&0.0587&0.0405&0.0951\\
$t$-ratio&15.87&17.21&4.75&4.87&3.24&1.02&19.85&13.44
    \end{tabular}}
    \caption{\textcolor{black}{Point estimates of the AR-HMM parameters at the end of the training sample under different initializations. The baseline case refers to the initialization procedure described in part (1) of Remark \ref{remark:implementationFilters}. The ``mean'' and ``std. dev.'' rows summarize parameter estimates from 100 runs of the filtering algorithm with random initializations, as described in Remark \ref{remark:stabilityEstimates}.}}
    \label{tab:estimatesFiltering}
\end{table}

\begin{remark}[Robustness of the estimates to filtering initialization]\label{remark:stabilityEstimates}
\textcolor{black}{To assess the stability of the parameter estimates with respect to the initialization of the filtering algorithm, we consider the two-state HMM specification and repeat the filter-based EM algorithm under randomized initial conditions. 
This procedure also serves to address potential local-maxima issues inherent in EM estimation (see, e.g., \cite[Chapter 3]{McLachlanKrishnan}).
We perform 100 independent runs of the algorithm. The transition probabilities $\pi_{11}$ and $\pi_{22}$ are independently drawn  from the interval $(0.5,1)$ in order to enforce a minimum degree of regime persistence in the hidden Markov chain. The initial guesses for the regime-dependent AR parameters are anchored to the preliminary OLS estimates obtained as in part (1) of Remark \ref{remark:implementationFilters}. For each parameter, lower and upper bounds are drawn uniformly from $(0.5,0.95)$ and $(1.05,1.5)$ times the corresponding OLS estimate, respectively. States are consistently labeled so that State 1 is associated with the larger value of $\gamma$. 
The results reported in Table \ref{tab:estimatesFiltering} indicate a remarkable stability in the estimated transition probabilities and limited variability in most regime-dependent parameters\footnote{In Table \ref{tab:estimatesFiltering}, the $t$-ratio (mean divided by standard deviation) provides a quantification of estimation uncertainty.}. For the majority of parameters, the mean estimate across initializations substantially exceeds the corresponding standard deviation, indicating limited sensitivity to initialization. An exception is the long-run mean in the second regime, whose estimate is close to zero. As documented in Remark \ref{remark:stabilityPerformances}, these mild instabilities have negligible impact on the performance of the statistical arbitrage strategies, which are primarily driven by opening signals and are therefore less sensitive to small deviations in the model parameters.
This indicates that the economically relevant outcomes of the model are robust to initialization and not driven by convergence to a specific local optimum.}
\end{remark}

\subsection{Testing the strategies}\label{subsec:TestingTheStrategies}
We now analyze over the test sample (from $t_B=$ 07/01/2022 until $T=$ 06/30/2023) the five statistical arbitrage strategies described in Section \ref{subsec:ParisTradingStrategies}: the plain Vanilla (PV), the probability interval (ProbI), the prediction interval (PredI), the realized increment (RI), the predicted increment (PI).  
%along with the buy-and-hold S\&P500 and the crude oil ETF strategies. 
\textcolor{black}{For all strategies except the PV benchmark, we consider several standard values of the bandwidth level $\alpha$}. As a comparison, we also include the buy-and-hold S\&P500 and the crude oil ETF strategies.

\begin{figure}
    \centering
    \includegraphics[width=\textwidth]{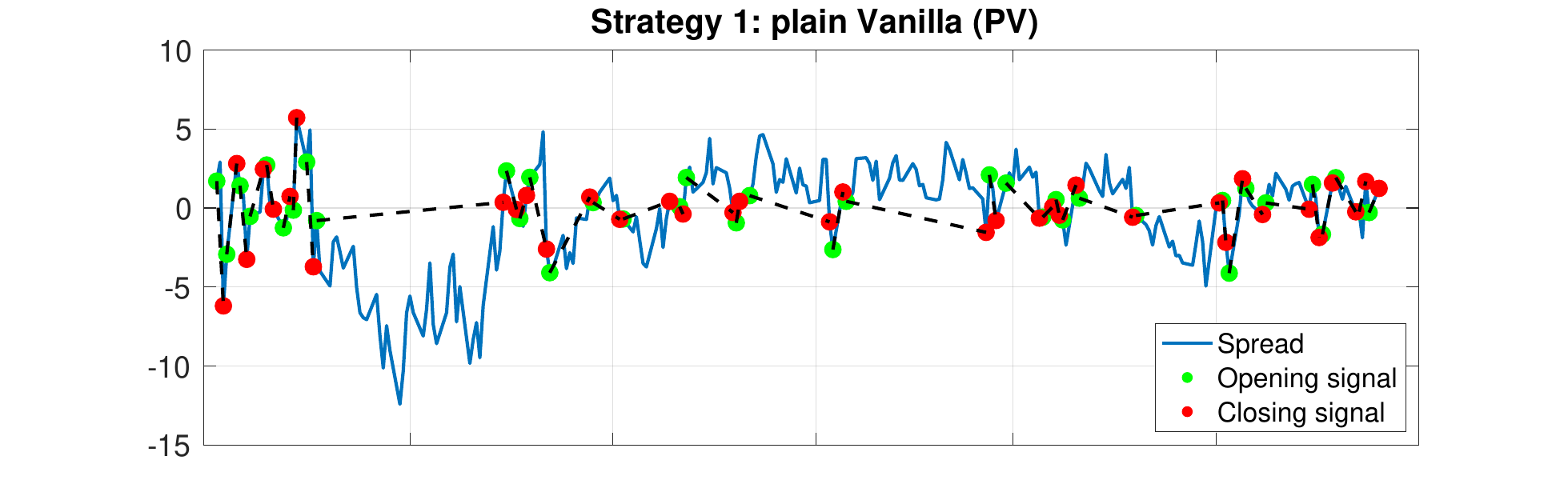}\\
    \vspace*{-13pt}
    \includegraphics[width=\textwidth]{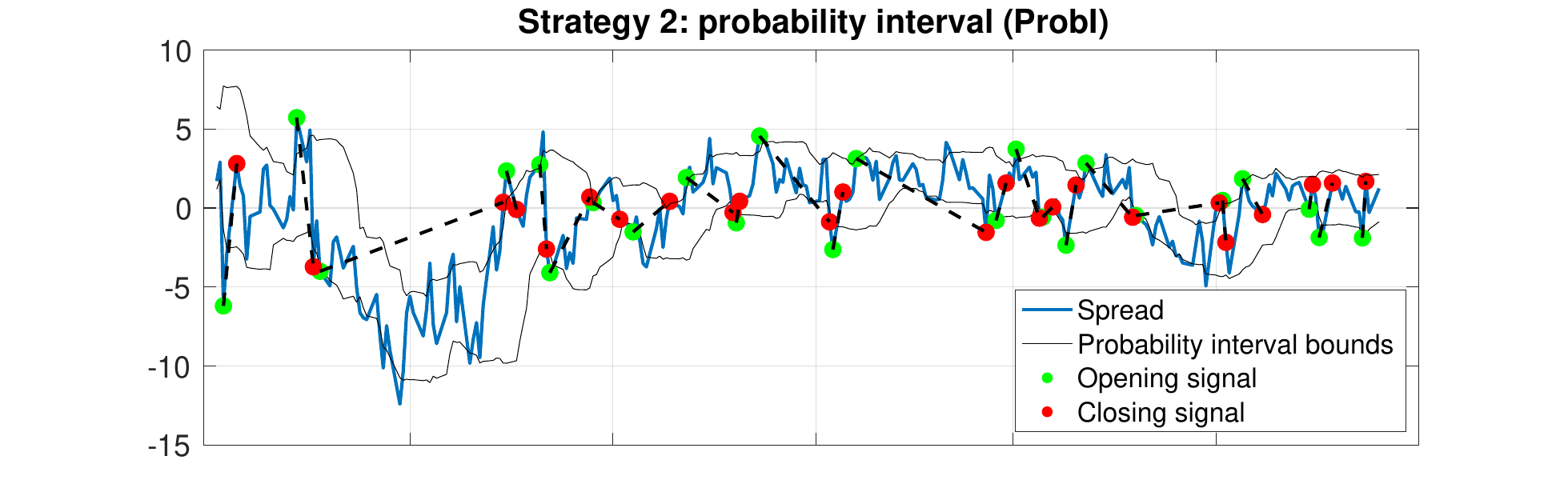}\\
    \vspace*{-13pt}
    \includegraphics[width=\textwidth]{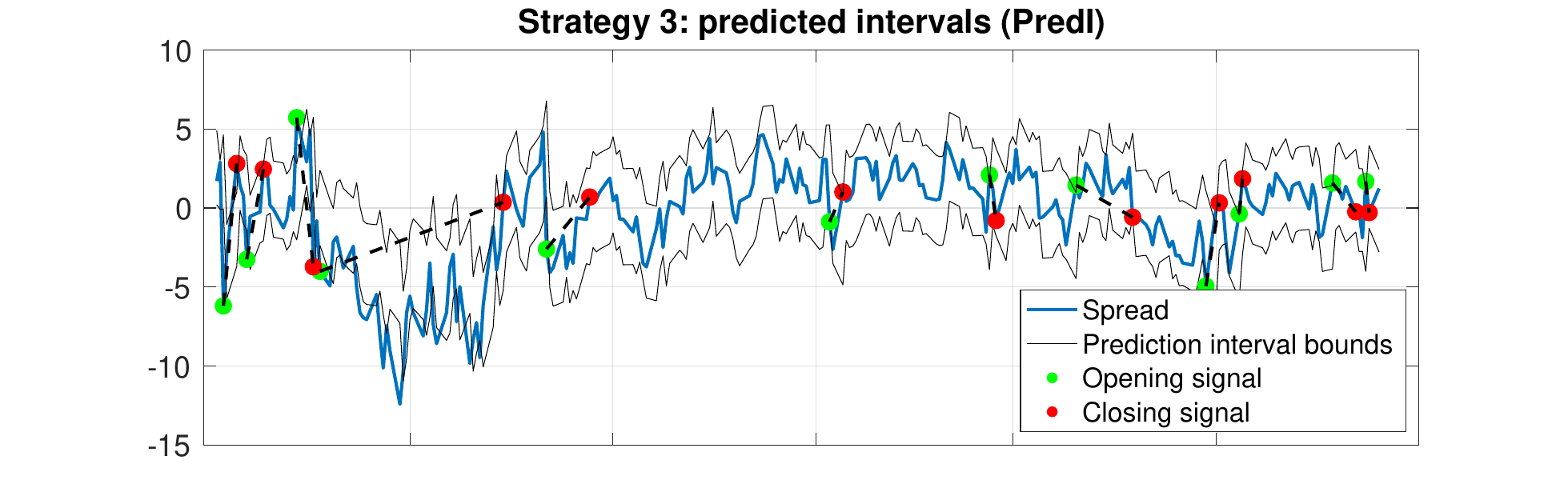}\\
    \vspace*{-13pt}
    \includegraphics[width=\textwidth]{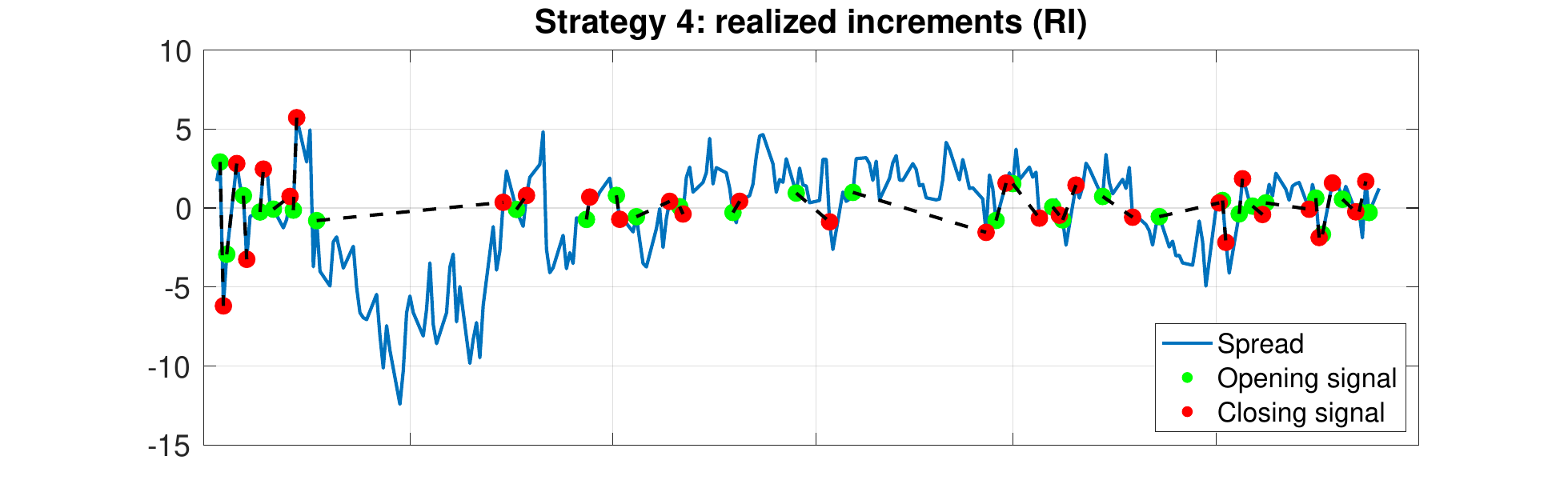}\\
    \vspace*{-13pt}
    \includegraphics[width=\textwidth]{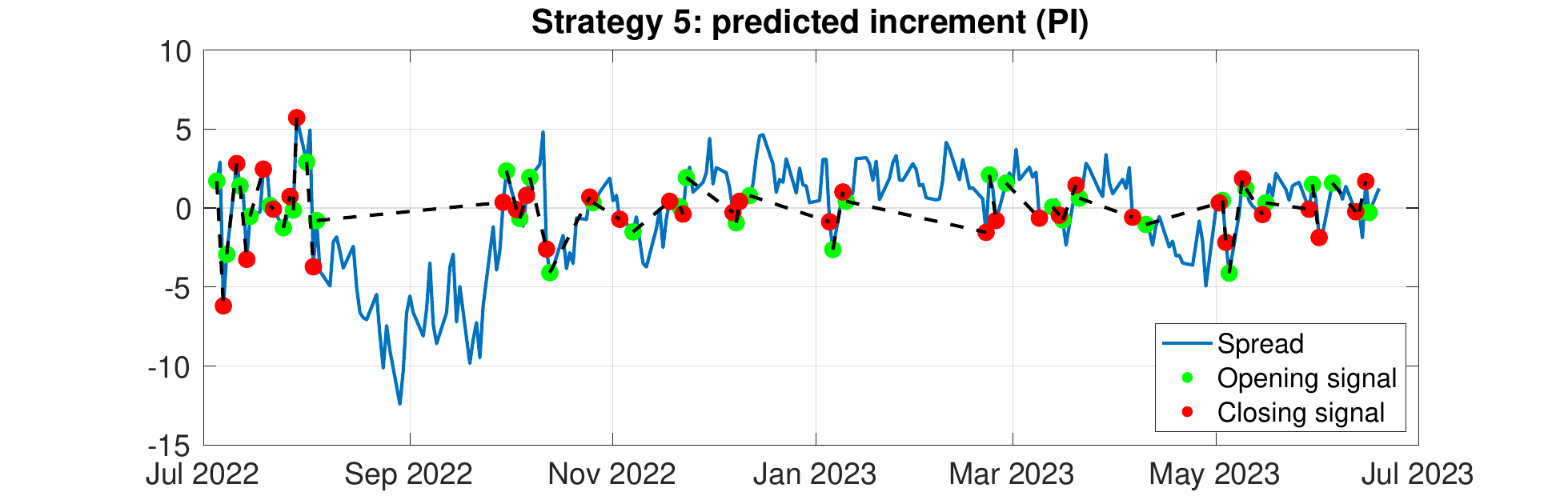}
    \caption{Opening/closing signals of the statistical arbitrage strategies described in Section \ref{subsec:ParisTradingStrategies}, with bandwith level $\alpha = 0.20$, over the test sample described in Section \ref{subsec:AnIllustrativeExample}.}
    \label{fig:Signals}
\end{figure}

Figure \ref{fig:Signals} illustrates how the trading signals differ across the five statistical arbitrage strategies, for the baseline value $\alpha = 0.20$ of the bandwidth level.
We immediately notice that the number of trades over the test sample varies significantly across the different strategies. Excluding the PV strategy that entails always an open position as soon as $S_t \neq 0$, we document a minimum of twelve trades with the PredI strategy and a maximum of thirty-five trades with the PI strategy. Moreover, also the length of the open positions varies considerably: indeed, we observe trading positions that are opened on a given day and closed on the following day together with positions that remain open for several months.

\begin{table}[]
    \centering
    \begin{small}
    \textcolor{black}{
    \begin{tabular}{cccccccccc}
&PV&ProbI&PredI&RI&PI&Ex. S\&P&Ex. ETF&\\
&&\multicolumn{4}{c}{$\alpha = 0.20$}&&&\\ \cline{2-6}
$R$&-0.0091&0.1043&0.1518&-0.0244&0.0230&0.1174&-0.2847&\\
$SR$&-0.0025&0.8335$^{*}$&1.1792$^{*}$&-0.1508&0.2359&0.6729&-0.7436&\\
$p$-val. SR test&0.5125&0.0765&0.0615&0.635&0.3790&0.2360&0.8710&\\
$N_{\mathrm{TW}}$&37&24&12&30&35&&&\\
mean $R^{\mathrm{TW}}$&-0.0002&0.0042$^{**}$&0.0120$^{**}$&-0.0008&0.0007&&&\\
$p$-val. $R^{\mathrm{TW}}$& 0.9053&0.0251&0.0467&0.6547&0.7013&&&\\
\% $R^{\mathrm{TW}}_j>0$&0.5135&0.6667&0.8333&0.4000&0.5714&&&\\ \cline{2-6}
$p$-val. WRC&&0.0650&\multicolumn{3}{l}{best from WRC: PredI$^{*}$}&&&\\
&&&&&&&&
\end{tabular}
\begin{tabular}{cccccccccc}
&ProbI&PredI&RI&PI&&ProbI&PredI&RI&PI\\
&\multicolumn{4}{c}{$\alpha = 0.30$}&&\multicolumn{4}{c}{$\alpha = 0.25$}\\ \cline{2-5} \cline{7-10}
$R$&0.1055&0.1394&-0.0471&0.0174&&0.1043&0.1668&-0.0432&0.0243\\
$SR$&0.8151$^{*}$&1.1010$^{*}$&-0.3137&0.1955&&0.8335$^{*}$&1.2562$^{**}$&-0.2909&0.2458\\
$p$-val. SR test&0.0795&0.0605&0.7270&0.3995&&0.0765&0.0360&0.7170&0.3715\\
$N_{\mathrm{TW}}$&26&18&35&36&&24&15&33&35\\
mean $R^{\mathrm{TW}}$&0.0039$^{**}$&0.0073$^{**}$&-0.0013&0.0005&&0.0042$^{**}$&0.0104$^{***}$&-0.0013&0.0007\\
$p$-val. $R^{\mathrm{TW}}$&0.0496&0.0202&0.4320&0.7639&&0.0251&0.0075&0.4702&0.6852\\
\% $R^{\mathrm{TW}}_j>0$&0.6538&0.8333&0.4000&0.5556&&0.6667&0.8667&0.3939&0.5714\\ \cline{2-5} \cline{7-10}
$p$-val. WRC&0.0745&\multicolumn{3}{l}{best from WRC: PredI$^{*}$}&&0.0360&\multicolumn{3}{l}{best from WRC: PredI$^{**}$}\\
&&&&&&&&&\\
&\multicolumn{4}{c}{$\alpha = 0.15$}&&\multicolumn{4}{c}{$\alpha = 0.05$}\\ \cline{2-5} \cline{7-10}
$R$&0.0572&0.1632&0.0206&0.0416&&0.0559&0.0729&-0.0020&0.1994\\
$SR$&0.5171&1.3032$^{**}$&0.2373&0.3685&&0.7556&0.7244&-0.0319&1.3136$^{**}$\\
$p$-val. SR test&0.1520&0.0180&0.3670&0.2925&&0.1030&0.1250&0.5290&0.0135\\
$N_{\mathrm{TW}}$&18&10&24&34&&7&5&11&27\\
mean $R^{\mathrm{TW}}$&0.0031&0.0153$^{**}$&0.0009&0.0013&&0.0079&0.0142$^{*}$&-0.0001&0.0068$^{**}$\\
$p$-val. $R^{\mathrm{TW}}$&0.2251&0.0110&0.6727&0.4944&&0.1657&0.0551&0.9720&0.0136\\
\% $R^{\mathrm{TW}}_j>0$&0.6667&0.90000&0.5417&0.6176&&0.7143&1.0000&0.5455&0.7407\\ \cline{2-5} \cline{7-10}
$p$-val. WRC&0.0335&\multicolumn{3}{l}{best from WRC: PredI$^{**}$}&&0.0110&\multicolumn{3}{l}{best from WRC: PI$^{**}$}
\end{tabular}}
    \end{small}
    \caption{\textcolor{black}{Performances of the statistical arbitrage strategies for different bandwidth levels $\alpha$ over the test sample of Section \ref{subsec:AnIllustrativeExample} considering the Brent, the Shanghai and the WTI futures.
    Statistical significance is denoted by $^{*}$, $^{**}$ and $^{***}$, corresponding to the 10\%, 5\% and 1\% levels, respectively.}}
    \label{tab:Performances}
\end{table}

\textcolor{black}{Table \ref{tab:Performances} summarizes the performance of all the strategies introduced in Section \ref{subsec:ParisTradingStrategies}, considering  different values of the bandwidth level $\alpha$ and the transaction cost parameters reported in Table \ref{tab:transactionCosts}. Note that the PV, Ex. S\&P, and Ex. ETF strategies do not depend on $\alpha$: accordingly, their performance measures are reported only once in the table.}

\textcolor{black}{To assess the statistical significance of the performance of the proposed strategies, Table \ref{tab:Performances} reports, alongside the performance metrics considered in Section \ref{subsec:PerformanceMeasure}, the results of several statistical tests. For each statistical arbitrage strategy, the following quantities are reported:
\begin{itemize}
    \item the overall annualized return $R$ over the test sample, as defined in (\ref{eqn:DefinitionR});
    \item the annualized Sharpe Ratio $SR$, defined in (\ref{eqn:DefinitionSR});
    \item the $p$-value of the test\footnote{\textcolor{black}{The test is implemented using the fixed-size circular block bootstrap considered in \cite{LedoitWolf2008}, based on 2{,}000 bootstrap replications with 20 blocks each.}} for the statistical significance of the Sharpe Ratio, as proposed in \cite{LedoitWolf2008}, conducted against the null hypothesis $H_0: SR=0$, which is appropriate since the performance of all strategies is evaluated in terms of excess returns;
    \item the total number $N_{\mathrm{TW}}$ of completed trades over the entire test sample;
    \item the sample mean of the trading-window returns $R^{\mathrm{TW}}=\{R^{\mathrm{TW}}_1,\ldots,R^{\mathrm{TW}}_{N_{\mathrm{TW}}}\}$;
    \item the $p$-value of the standard $t$-test for the null hypothesis that the mean of $R^{\mathrm{TW}}$ is zero;
    \item the percentage of strictly positive trading-window returns in $R^{\mathrm{TW}}$;
    \item the $p$-value of the White Reality Check (WRC)\footnote{\textcolor{black}{The test is implemented using the stationary bootstrap of \cite{PolitisRomano1994}, with 2{,}000 bootstrap replications and an average block length of 20.}} proposed by \cite{White2000} and implemented following \cite{SullivanTimmermannWhite1999}. This test evaluates whether, within a set of strategies backtested on the same dataset, at least one strategy significantly outperforms both the others and a benchmark, here again taken to be a zero-return investment. Along with the WRC $p$-value, we also report the best-performing strategy in our sample.
\end{itemize}}

First of all, we notice that the ETF on the crude oil futures market performs poorly on the test sample. This is coherent with the global decline of crude oil futures prices over the test sample which is clear from Figure \ref{fig:futuresPrices_BSW}. In contrast, the S\&P performed relatively well, \textcolor{black}{but its Sharpe Ratio is not statistically significant at any conventional confidence level.}

\textcolor{black}{Overall, the results reported in Table~\ref{tab:Performances} indicate a clear advantage of the forward-looking strategies based on the proposed AR-HMM model, namely PredI and PI, over the model-free backward-looking strategies ProbI and RI. Across all considered values of the bandwidth level, PredI consistently yields the highest annualized returns and Sharpe ratios among the statistical arbitrage strategies, often attaining statistical significance according to the Sharpe Ratio test. The PI strategy also performs strongly, particularly for smaller values of $\alpha$, where it combines high returns with a limited number of trades and favorable trading-window statistics. In contrast, the backward-looking ProbI strategy exhibits positive but generally weaker performance, while the RI strategy fails to deliver consistent profitability and is never statistically significant. The superiority of the forward-looking approaches is further confirmed by the White Reality Check, which repeatedly identifies PredI (and, for the smallest value of $\alpha$, PI) as the best-performing strategy within the set, with rejection of the null hypothesis of no superior strategy at conventional significance levels. Taken together, these results suggest that explicitly modeling the joint dynamics of the spread and its latent regimes yields economically and statistically meaningful gains relative to purely distributional backward-looking alternatives.}

Finally, we observe that the magnitude of these returns is consistent with the results in \cite{HainHessUhrigHomburg2018}, that document that this kind of statistical arbitrage strategies involving commodity futures yield excess returns of about 6\%-8\% annually.

\begin{remark}
\textcolor{black}{It is worth noting that, for all statistical arbitrage strategies tested in our study, the series of daily returns include extended periods in which no position is open and returns are therefore equal zero. 
The resulting dilution of the unconditional daily mean may attenuate the Sharpe ratio relative to the profitability realized during active trading windows, potentially reducing the power of statistical Sharpe ratio tests based on daily returns. This issue further motivates the interest of performing a complementary analysis based on trading-window returns, as introduced in Section \ref{subsec:PerformanceMeasure}.}
\end{remark}

\begin{remark}[Robustness of trading strategy performance to filtering initialization]\label{remark:stabilityPerformances}
\textcolor{black}{Table \ref{tab:RobustnessPerformancesFiltering} provides evidence on the robustness of the performance of the PredI and PI strategies with respect to the initialization of the filtering algorithm. For conciseness, we report the annualized return $R$, the Sharpe ratio $SR$, and the mean trading-window return $R^{\mathrm{TW}}$. Across 100 runs of the filter-based EM algorithm with randomized initial conditions as described in Remark \ref{remark:stabilityEstimates}, both strategies exhibit limited dispersion in all reported performance measures.
The standard deviations of $R$, $SR$ and $R^{\mathrm{TW}}$ are small relative to their corresponding means, resulting in consistently large $t$-ratios. This pattern indicates that the observed profitability is not driven by specific initialization choices or by convergence to isolated local optima of the likelihood, but instead reflects stable features of the underlying spread modeling and filtering framework.}
\end{remark}

\begin{table}[]
    \centering
    \textcolor{black}{
    \begin{tabular}{c|ccc|ccc}
         &\multicolumn{3}{c|}{PredI}&\multicolumn{3}{c}{PI}\\
&$R$&$SR$&mean $R^{\mathrm{TW}}$&$R$&$SR$&mean $R^{\mathrm{TW}}$\\ \hline
baseline&0.1518&1.1792&0.012&0.023&0.2359&0.0007\\
mean&0.1679&1.2718&0.0125&0.0374&0.3406&0.0012\\
std. dev.&0.0349&0.254&0.0033&0.0094&0.0706&0.0005\\
$t$-ratio&4.808&5.008&3.8159&4.0001&4.8235&2.484\\
    \end{tabular}}
    \caption{\textcolor{black}{Robustness of the performance of the PredI and PI strategies with respect to the initialization of the filtering algorithm. Reported means and standard deviations are computed from 100 runs of the algorithm with random initializations, as described in Remark \ref{remark:stabilityEstimates}.}}
    \label{tab:RobustnessPerformancesFiltering}
\end{table}

\subsubsection{Risk analysis}\label{subsec:VaRAnalysis}
Statistical arbitrage strategies might involve a considerable financial risk. In particular, when opening a position, the investor does not know when the futures prices will realign or whether convergence will ever be reached. As a consequence, the investor might find herself stuck in a costly position to maintain (due to margin calls) and that does not generate  profits for a prolonged period of time. In this section, we assess the riskiness of the statistical arbitrage strategies by means of a bootstrap-based analysis.
\textcolor{black}{We conduct the analysis using the baseline bandwidth level $\alpha = 0.20$, an intermediate specification within the range considered above that balances trading frequency and transaction costs, as discussed in Section \ref{subsec:ParisTradingStrategies}.}

To provide a comprehensive assessment of the risk profile of the trading strategies described in Section~\ref{subsec:ParisTradingStrategies}, and to overcome the limitations of the Sharpe ratio, which depends solely on the first two moments of the return distribution, we adopt a resampling-based approach. Specifically, we employ the stationary bootstrap of \cite{PolitisRomano1994} with 2{,}000 replications and an expected block length of 20 to account for serial dependence in daily returns.
Using this procedure, we compute out-of-sample risk measures at  daily frequency, including the left-tail Value-at-Risk (VaR), the corresponding Expected Shortfall (ES) and the Maximum Drawdown (MDD). In addition, the bootstrap resampling scheme allows us to recover the empirical distribution of annual returns $R$. This enables a full characterization of the distributional properties of strategy performance beyond location and scale measures.

\begin{table}
    \centering
    \begin{tabular}{c|ccccccc}
         &PV&ProbI&PredI&RI&PI&Ex. S\&P&Ex. ETF\\ \hline
VaR$_{99\%}$&-2.5078\%&-2.5078\%&-2.5078\%&-2.6663\%&-2.5078\%&-2.8106\%&-5.3009\% \\
VaR$_{95\%}$&-1.5197\%&-1.3472\%&-1.3178\%&-1.4422\%&-1.5197\%&-1.7337\%&-3.8152\% \\
VaR$_{90\%}$&-1.1033\%&-1.0294\%&-0.8052\%&-1.0294\%&-1.1033\%&-1.3019\%&-3.1380\% \\
&&&&&&&\\
ES$_{99\%}$&-2.7297\%&-2.7297\%&-2.7297\%&-2.8093\%&-2.7297\%&-3.5073\%&-6.9106\% \\
ES$_{95\%}$&-1.8981\%&-1.8324\%&-1.8155\%&-1.896\%&-1.8981\%&-2.435\%&-5.0296\% \\
ES$_{90\%}$&-1.5848\%&-1.5259\%&-1.4400\%&-1.5728\%&-1.5848\%&-1.9961\%&-4.2791\% \\
&&&&&&&\\
MDD&-10.1668\%&-6.3788\%&-5.7282\%&-10.7311\%&-9.3580\%&-15.5447\%&-39.4747\%
    \end{tabular}
    \caption{Bootstrapped estimates of the daily VaRs/ESs/MDD of daily returns over the test sample of Section \ref{subsec:AnIllustrativeExample} including the Brent, the Shanghai and the WTI futures contracts.}
    \label{tab:VaRShangai}
\end{table}
The results reported in Table \ref{tab:VaRShangai} indicate that the proposed strategies exhibit substantially lower tail risk and drawdown relative to the excess return benchmarks, in particular when compared to the Ex. ETF strategy, which displays  larger losses and significantly deeper drawdowns. By contrast, the proposed statistical arbitrage strategies tend to generate less volatile returns. This aligns with the trading signals shown in Figure \ref{fig:Signals}, where, apart from one long position shortly after the start of the test period, most positions are held for relatively short durations, allowing for quick convergence and the realization of positive profits. While mild differences across strategies emerge at conventional levels, at the 99\% level the scarcity of extreme observations limits the empirical discrimination of downside risk measures.

\begin{figure}
    \centering
    \includegraphics[width=0.48\textwidth]{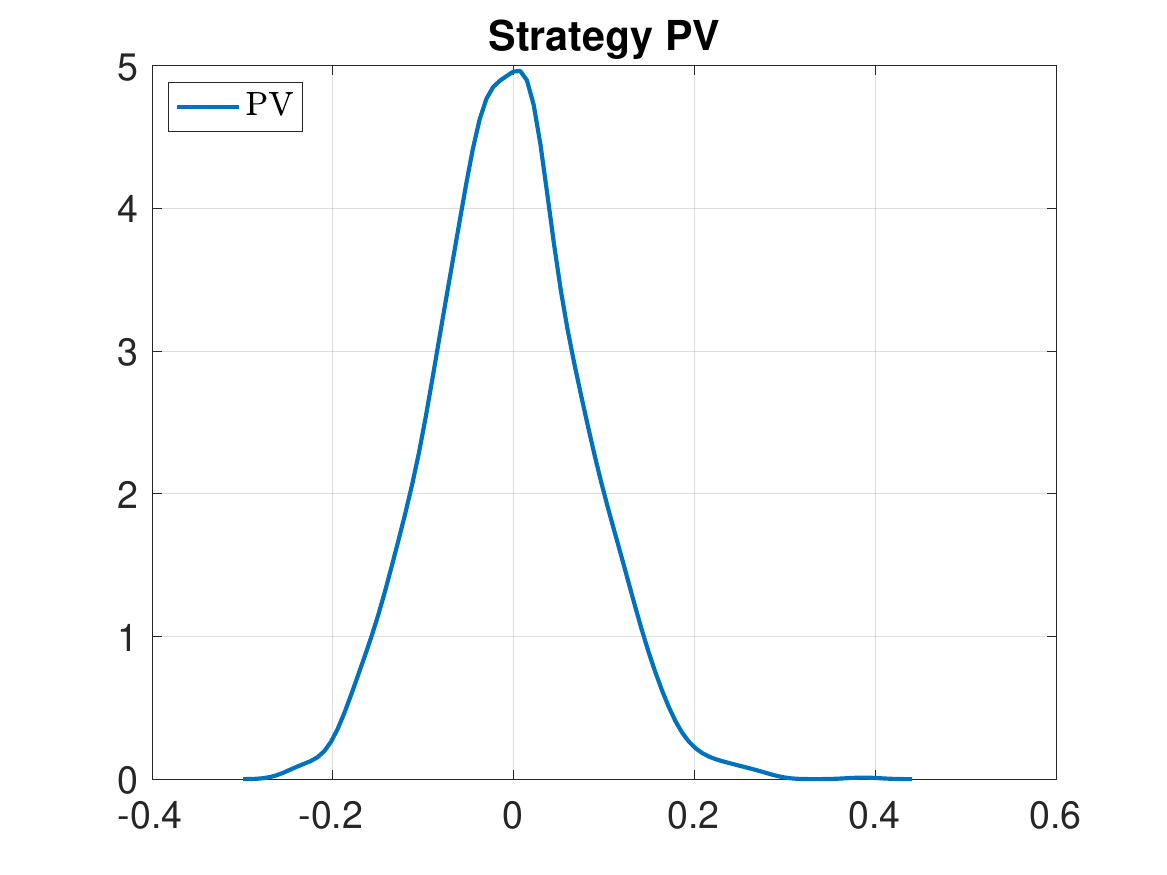}
    \includegraphics[width=0.48\textwidth]{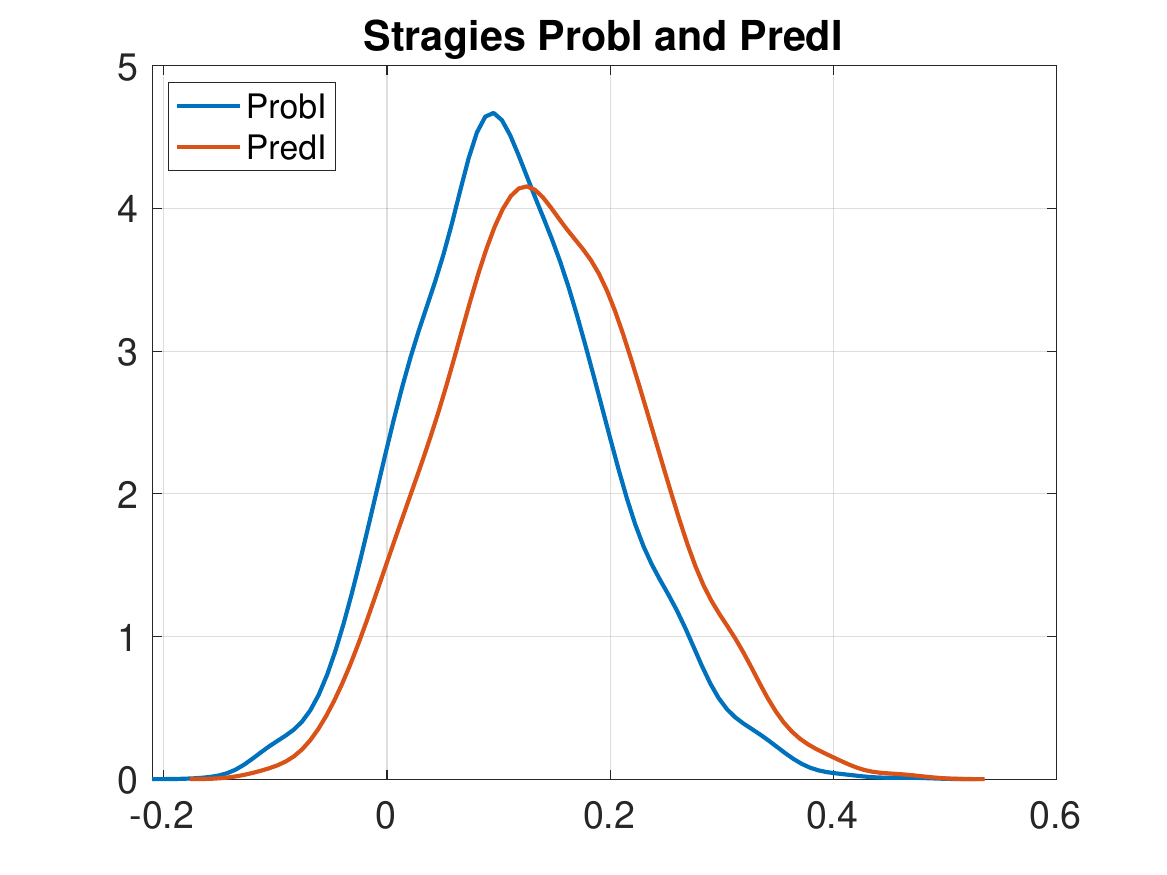}\\
    \includegraphics[width=0.48\textwidth]{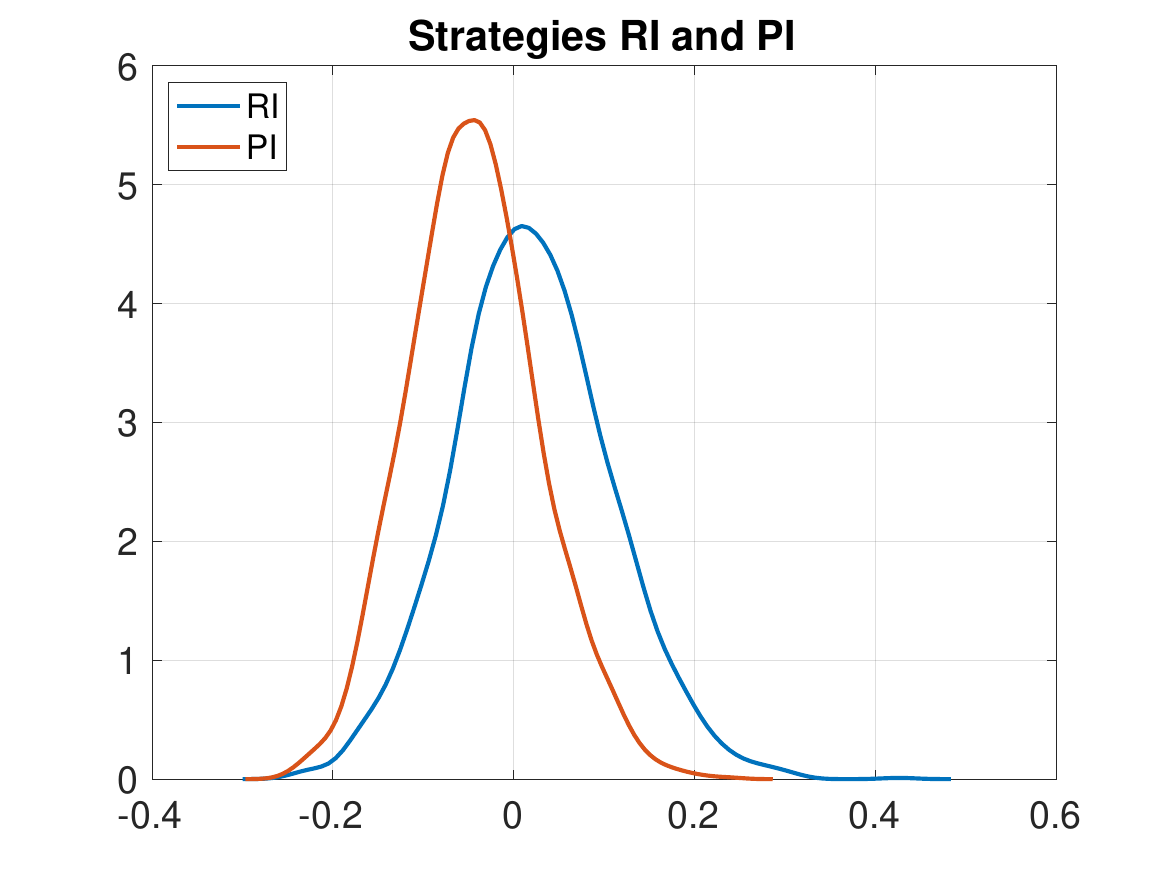}
    \caption{Bootstrapped KDEs of annual returns over the test sample of Section \ref{subsec:AnIllustrativeExample} including the Brent, the Shanghai and the WTI futures contracts.}
    \label{fig:KDEs_Shangai}
\end{figure}

Figure \ref{fig:KDEs_Shangai} reports the bootstrap-based kernel density estimates of the annualized returns of the five strategies introduced in Section \ref{subsec:ParisTradingStrategies} over the test sample. 
The dispersion and location of the distributions differ across strategy classes. In particular, the plain vanilla (PV) strategy exhibits a comparatively symmetric distribution concentrated around its central tendency.
We can observe that for the ProbI and PredI strategies the estimated densities are centered around positive returns.
The density of the PredI strategy is moderately shifted to the right relative to that of its backward-looking counterpart, ProbI, suggesting a tendency toward higher returns. By contrast, the distributions of the RI and PI strategies display noticeable differences in shape and dispersion, with neither strategy  dominating the other across the support of the distribution.

\subsection{Choice of the underlying futures contracts}\label{subsec:DifferentUnderlyings}
In order to assess the relevance of the results reported in the previous subsection, we test the same statistical arbitrage strategies using different choices of futures contracts. First, we follow the traditional approach to pairs trading by using strategies that involve only two futures contracts instead of three. Then, we consider an alternative triplet of securities, replacing the recently introduced Shanghai futures with the Dubai crude oil futures, a well-established benchmark for crude oil. Finally, we also consider a strategy involving all the four futures contracts: Brent, Dubai, Shanghai, and WTI.

The results of this analysis are reported in Table \ref{tab:ProfitsCouples}\textcolor{black}{, where, for the sake of brevity, we fix the bandwidth level $\alpha$ at $0.20$ and omit the $p$-values associated with the \cite{LedoitWolf2008} test for the Sharpe ratio and the standard $t$-test for the mean trading-window returns, indicating statistical significance using the conventional notation instead.} For comparison purposes, the first row replicates the data from Table \ref{tab:Performances}, while the performances of the benchmark passive strategies, S\&P ($R$: 11.74\%, $SR$: 0.6729) and ETF ($R$: -28.47\%, $SR$: -0.7436), are omitted. Transaction costs specific to each futures contract are again included, as specified in Table \ref{tab:transactionCosts}. Interestingly, when evaluating potential pairs of securities for our strategies, some combinations did not exhibit statistical evidence of cointegration\footnote{As explained in Section \ref{subsec:UnitRootsVARCointegration}, cointegration is initially tested using the Johansen methodology. When this test did not detect cointegration, we also applied the Engle-Granger two-step procedure, considering both choices of dependent variable. Even with this alternative method, no evidence of cointegration was found.}, at least over our training sample (these pairs include WTI-Dubai, Shanghai-WTI and, perhaps unexpectedly, Brent-WTI). This suggests that including at least three contracts in the strategy can be advantageous when standard cointegration does not hold between individual pairs.

\textcolor{black}{As shown in Table \ref{tab:ProfitsCouples}, the statistical arbitrage strategies based on our modeling framework, namely the PredI and PI strategies, exhibit consistently strong performance across the different sets of securities considered. According to the White Reality Check of \cite{SullivanTimmermannWhite1999}, one of these two strategies is always identified as the best-performing alternative. Moreover, the majority of the reported performance measures are statistically significant, with the PredI strategy in particular delivering robust and economically meaningful results across all configurations.}

\begin{table}
    \centering
    \textcolor{black}{
    \begin{tabular}{cccccc}
        &PV&ProbI&PredI&RI&PI\\
&&&&&\\
&\multicolumn{5}{c}{Brent - Shanghai - WTI}\\ \hline
$R$&-0.0091&0.1043&0.1518&-0.0244&0.023\\
$SR$&-0.0025&0.8335$^{*}$&1.1792$^{*}$&-0.1508&0.2359\\
mean $R^{\mathrm{TW}}$&-0.0002&0.0042$^{**}$&0.012$^{**}$&-0.0008&0.0007\\
$p$-val. WRC&0.065&\multicolumn{4}{l}{best from WRC: PredI$^{*}$}\\
&&&&&\\
&\multicolumn{5}{c}{Brent - Shanghai}\\ \hline
$R$&0.1316&0.104&0.3621&0.0359&0.1711\\
$SR$&0.7561$^{*}$&0.6382&1.7738$^{***}$&0.2856&0.9302$^{*}$\\
mean $R^{\mathrm{TW}}$&0.0041&0.0044&0.0225$^{***}$&0.0014&0.0054\\
$p$-val. WRC&0.011&\multicolumn{4}{l}{best from WRC: PredI$^{**}$}\\
&&&&&\\
&\multicolumn{5}{c}{Brent - Dubai}\\ \hline
$R$&0.1478&0.099&0.1214&0.0692&0.1479\\
$SR$&1.8001$^{***}$&1.3823$^{**}$&1.5267$^{***}$&1.1975$^{*}$&1.8156$^{***}$\\
mean $R^{\mathrm{TW}}$&0.005$^{***}$&0.0056$^{***}$&0.007$^{***}$&0.0032$^{**}$&0.0058$^{***}$\\
$p$-val. WRC&0.0035&\multicolumn{4}{l}{best from WRC: PI$^{***}$}\\
&&&&&\\
&\multicolumn{5}{c}{Brent - Shanghai - Dubai}\\ \hline
$R$&0.1014&0.0329&0.0692&0.0014&0.1129\\
$SR$&1.229$^{**}$&0.5189&0.915$^{**}$&0.053&1.3569$^{**}$\\
mean $R^{\mathrm{TW}}$&0.003$^{**}$&0.0019&0.0056$^{**}$&0.0001&0.0041$^{***}$\\
$p$-val. WRC&0.0135&\multicolumn{4}{l}{best from WRC: PI$^{**}$}\\
&&&&&\\
&\multicolumn{5}{c}{Brent - WTI - Dubai}\\ \hline
$R$&0.113&0.069&0.122&0.0511&0.1163\\
$SR$&1.4469$^{**}$&1.007$^{*}$&1.5607$^{***}$&0.8929&1.4864$^{**}$\\
mean $R^{\mathrm{TW}}$&0.0045$^{**}$&0.0042$^{**}$&0.007$^{***}$&0.0026&0.0053$^{***}$\\
$p$-val. WRC&0.009&\multicolumn{4}{l}{best from WRC: PredI$^{***}$}\\
&&&&&\\
&\multicolumn{5}{c}{Brent - Shanghai - WTI - Dubai}\\ \hline
$R$&0.1206&0.0554&0.0636&0.0085&0.1343\\
$SR$&1.4311$^{***}$&0.7912$^{*}$&0.8691$^{*}$&0.1657&1.5781$^{***}$\\
mean $R^{\mathrm{TW}}$&0.0035$^{**}$&0.0032$^{*}$&0.0042&0.0004&0.0045$^{***}$\\
$p$-val. WRC&0.004&\multicolumn{4}{l}{best from WRC: PI$^{***}$}\\
    \end{tabular}}
    \caption{Performances over the test sample including different combinations of the Brent, the Dubai, the Shanghai and the WTI futures contracts.}
    \label{tab:ProfitsCouples}
\end{table}

\subsection{Robustness analysis}\label{subsec:validationOfTheResults}
To further assess the robustness of our findings, we now systematically repeat the analysis carried out in the previous sections for different values of the transaction costs parameters $c_i$'s, the starting date $t_0$ of the sample and the breaking date $t_B$ that separates the training sample from the test one.
\textcolor{black}{Similarly as in Section \ref{subsec:VaRAnalysis}, the analysis is performed using the baseline bandwidth level $\alpha = 0.20$, representing an balanced specification among the range of $\alpha$ considered in Section \ref{subsec:TestingTheStrategies}.}

\subsubsection{Sensitivity with respect to transaction costs}
\begin{figure}
    \centering
    \includegraphics[width=0.48\textwidth]{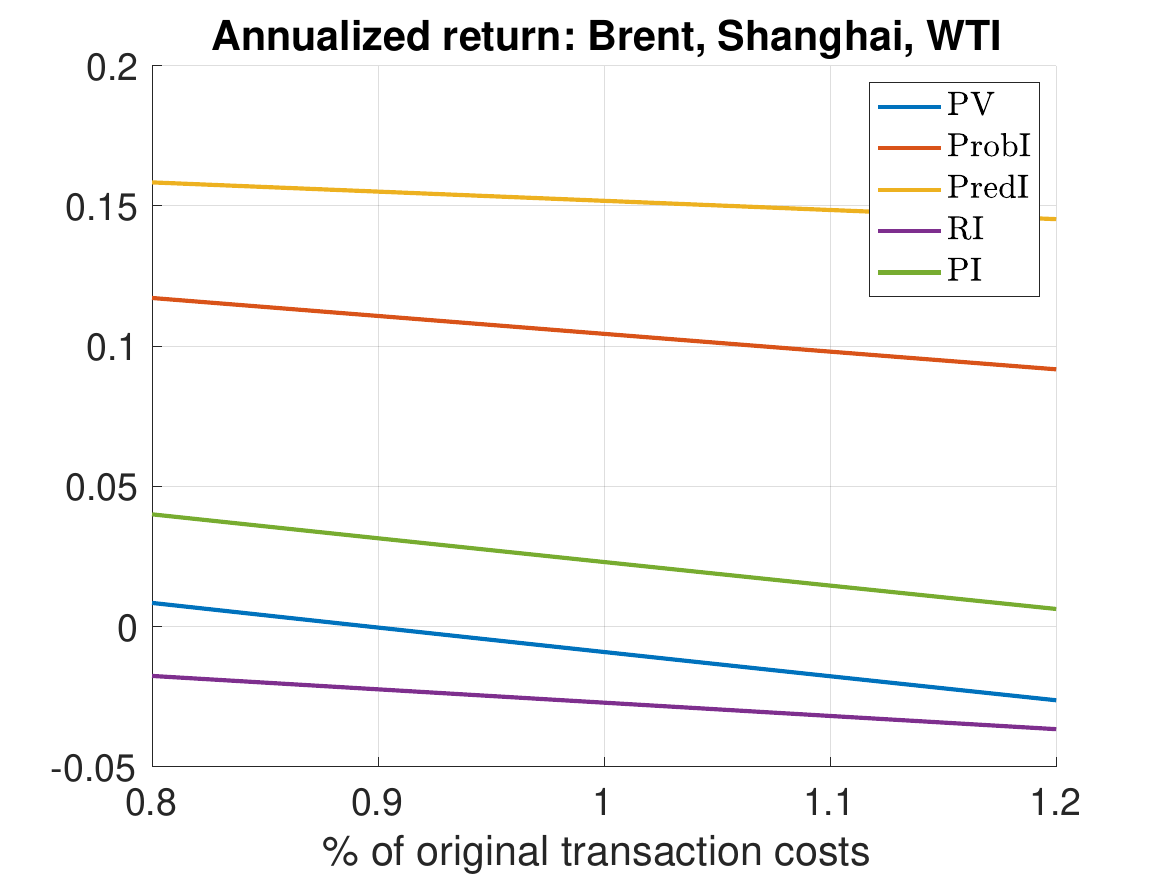}
    \includegraphics[width=0.48\textwidth]{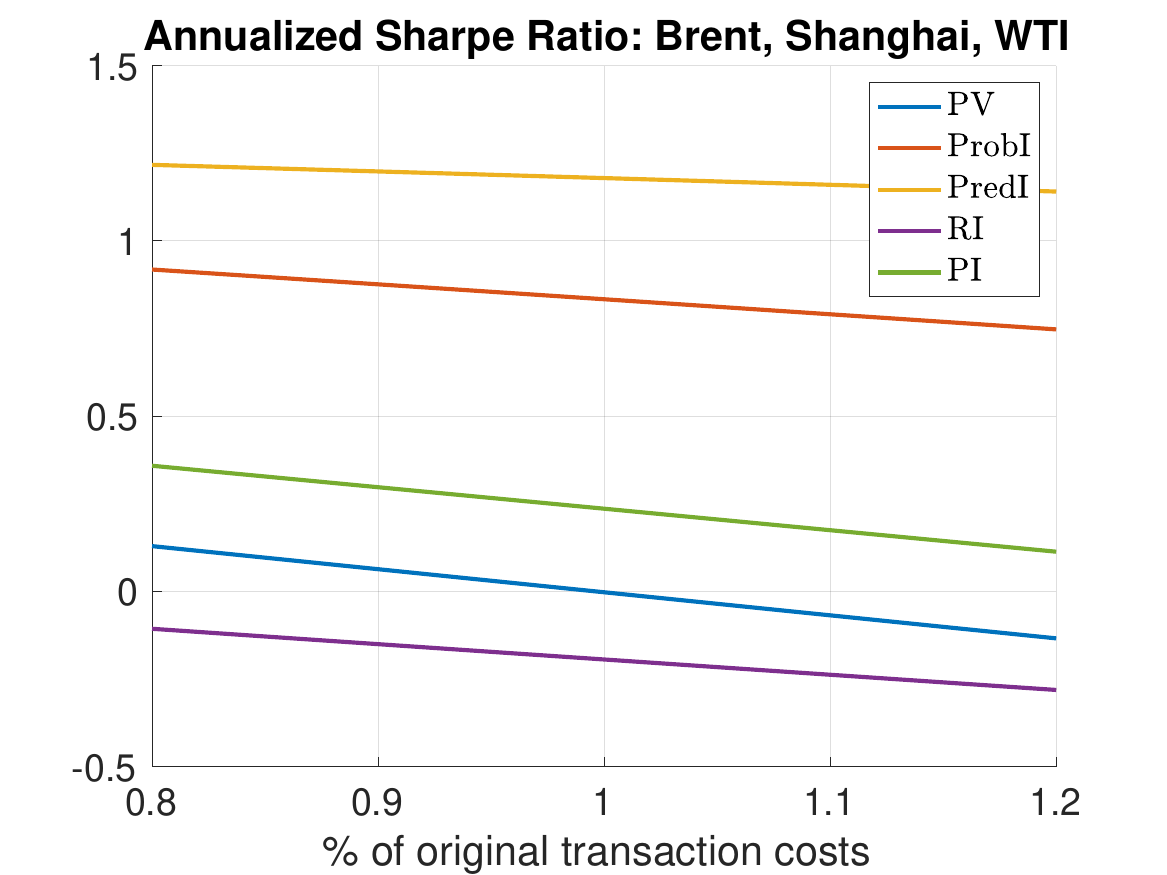}
    \caption{Performances of the strategies with respect to different percentages of the vector collecting the original transaction costs $c_i$ in Table \ref{tab:transactionCosts}; $t_0=$ 03/26/2018, $t_B=$ 07/01/2022.}
    \label{fig:Performances_BSW_c}
\end{figure}
Figure \ref{fig:Performances_BSW_c} displays the performances of the statistical arbitrage strategies as a function of a multiple of the original values of the transaction cost parameters in Table \ref{tab:transactionCosts}. Obviously, both the annualized return $R$ and the Sharpe ratio $SR$ are declining in $c$. The slope of this decline is proportional to the frequency of opening/closing positions: the performances of strategies PV, PredI and PI that prescribe the opening of several trades (see Figure \ref{fig:Signals}) are the most sensible to the $c$'s while strategies ProbI and RI are less affected by transaction costs. We find that, over the considered time period, statistical arbitrage strategies that include the Shanghai futures are quite robust with respect to the level of transaction costs.
It is also noteworthy that the strategies that make use of the AR-HMM model introduced in Section \ref{subsec:OUHMM}, namely the PredI and the PI strategies, turn out to be very profitable for small (but realistic) levels of transaction costs. In our view, this represents a further evidence of the potential advantages of relying on a stochastic model in the design of statistical arbitrage strategies.

\subsubsection{Sensitivity with respect to $t_0$}
\begin{figure}
    \centering
    \includegraphics[width=\textwidth]{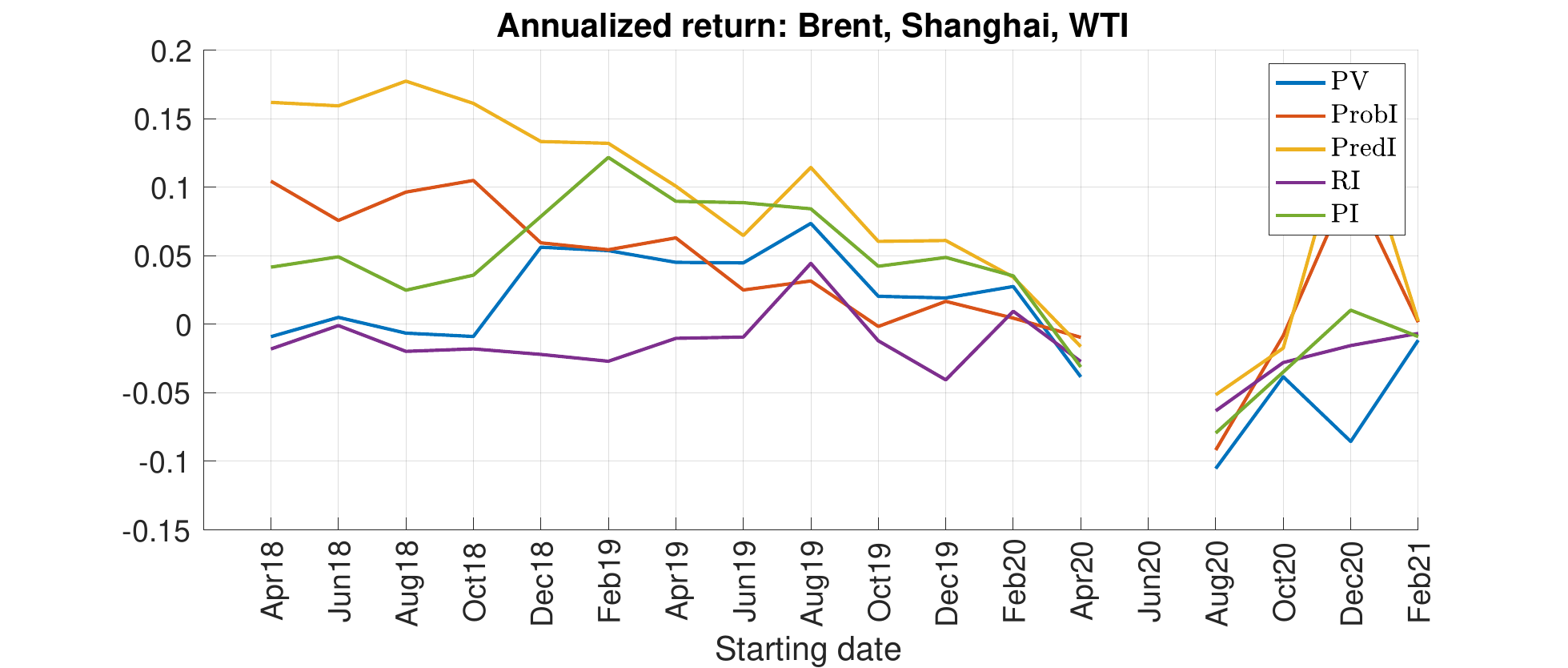}\\
    \vspace{-13pt}
    \includegraphics[width=\textwidth]{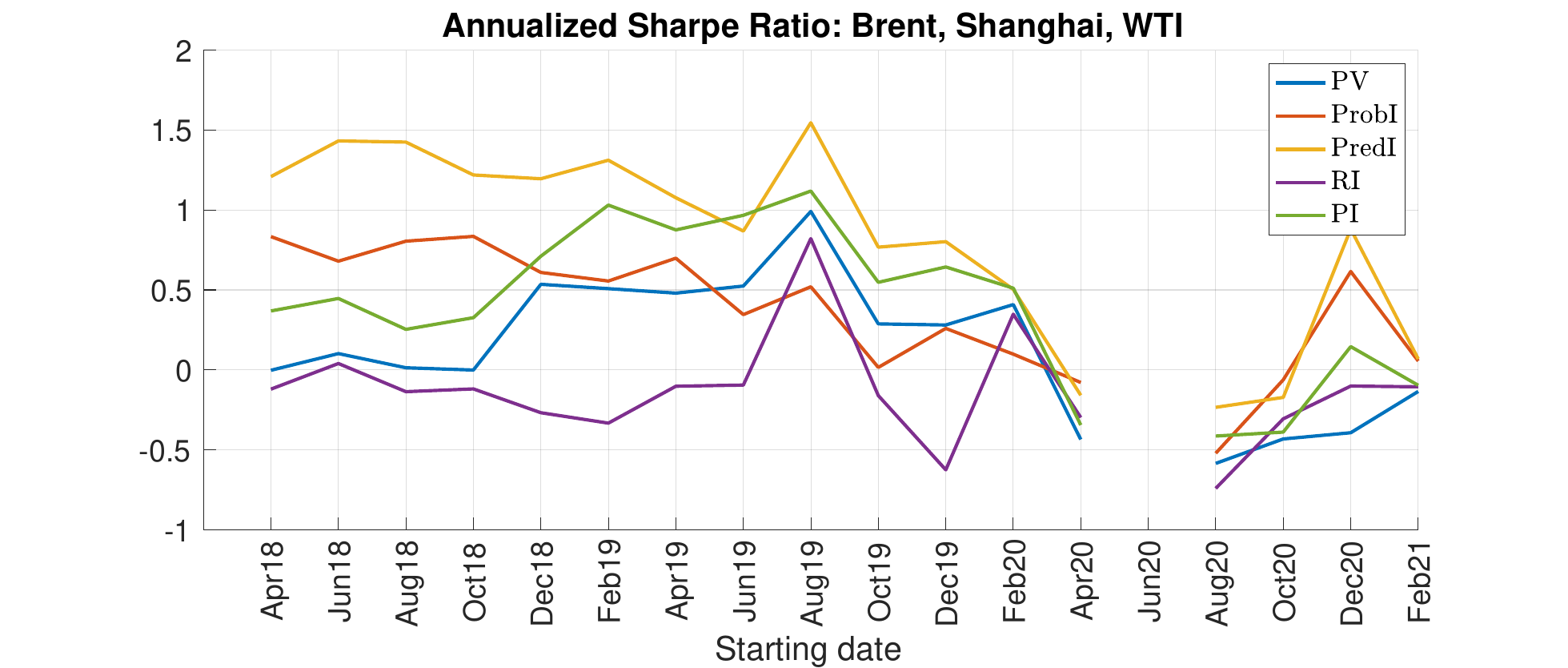}
    \caption{Performances of the strategies with respect to $t_0$; $t_B=$ 07/01/2022. Missing datapoints are due to the rejection of cointegration among the futures contract over the related time window.}
    \label{fig:Performances_BSW_t0}
\end{figure}
Figure \ref{fig:Performances_BSW_t0} depicts the performance of our strategies as a function of the starting date $t_0$ of the sample. Since the breaking date $t_B$ is fixed, if $t_0$ moves forward the number of observations in the training sample shrinks. 
This is expected to worsen the quality of the filter-based estimates derived according to the algorithm described in Section \ref{subsec:filter_EM}, due to the availability of a smaller number of observations.
As we can see from Figure \ref{fig:Performances_BSW_t0}, reducing the training sample weakens the cointegrating relationship among the time series and, as a consequence, lowers the overall performance of our strategies. 
This indicates that, in order to be profitable, our statistical arbitrage strategies require a minimum sample size for the model to be accurately estimated.
From Figure \ref{fig:Performances_BSW_t0} we can also notice that the dataset is significantly affected by the market turmoil of the COVID-19 pandemics. Given the rather recent introduction of the Shanghai futures, this issue cannot be avoided, since the available dataset is relatively small.

\subsubsection{Sensitivity with respect to $t_B$}
\begin{figure}
    \centering
    \includegraphics[width=\textwidth]{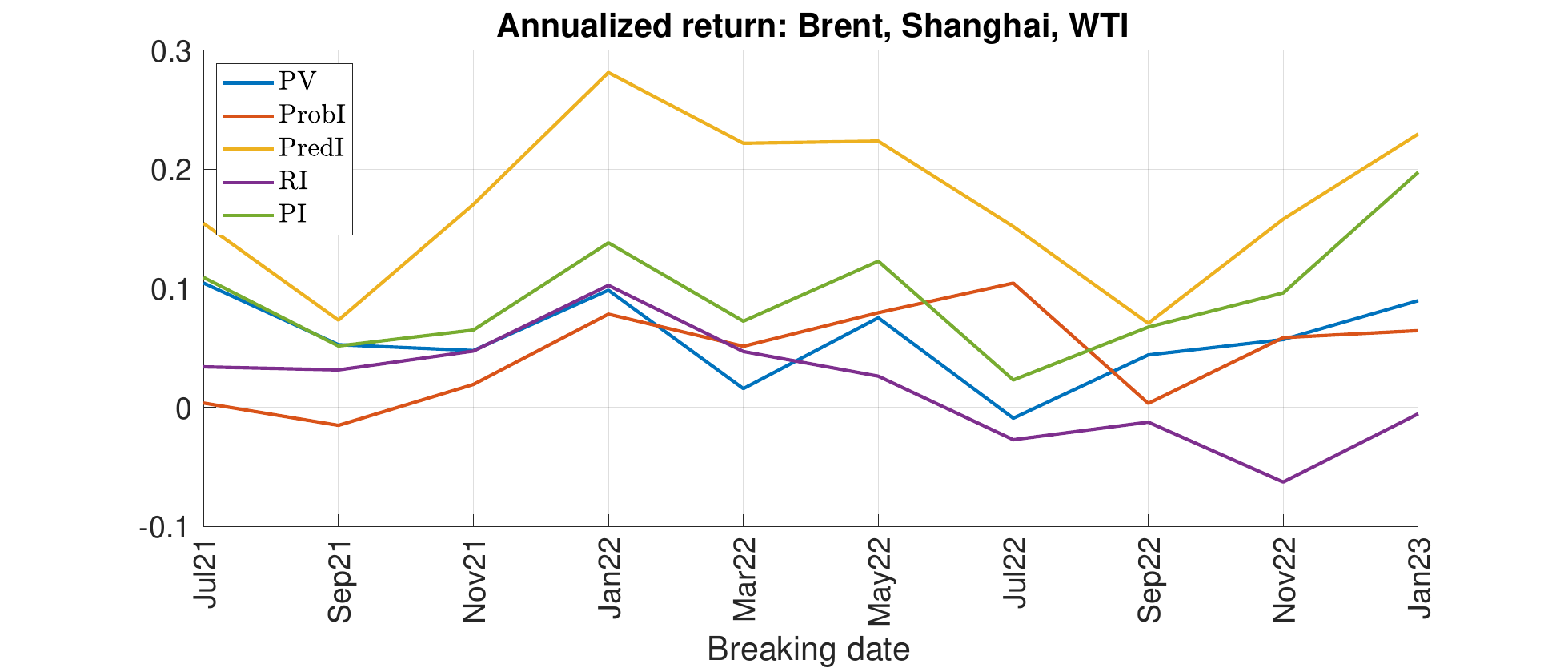}\\
    \vspace{-13pt}
    \includegraphics[width=\textwidth]{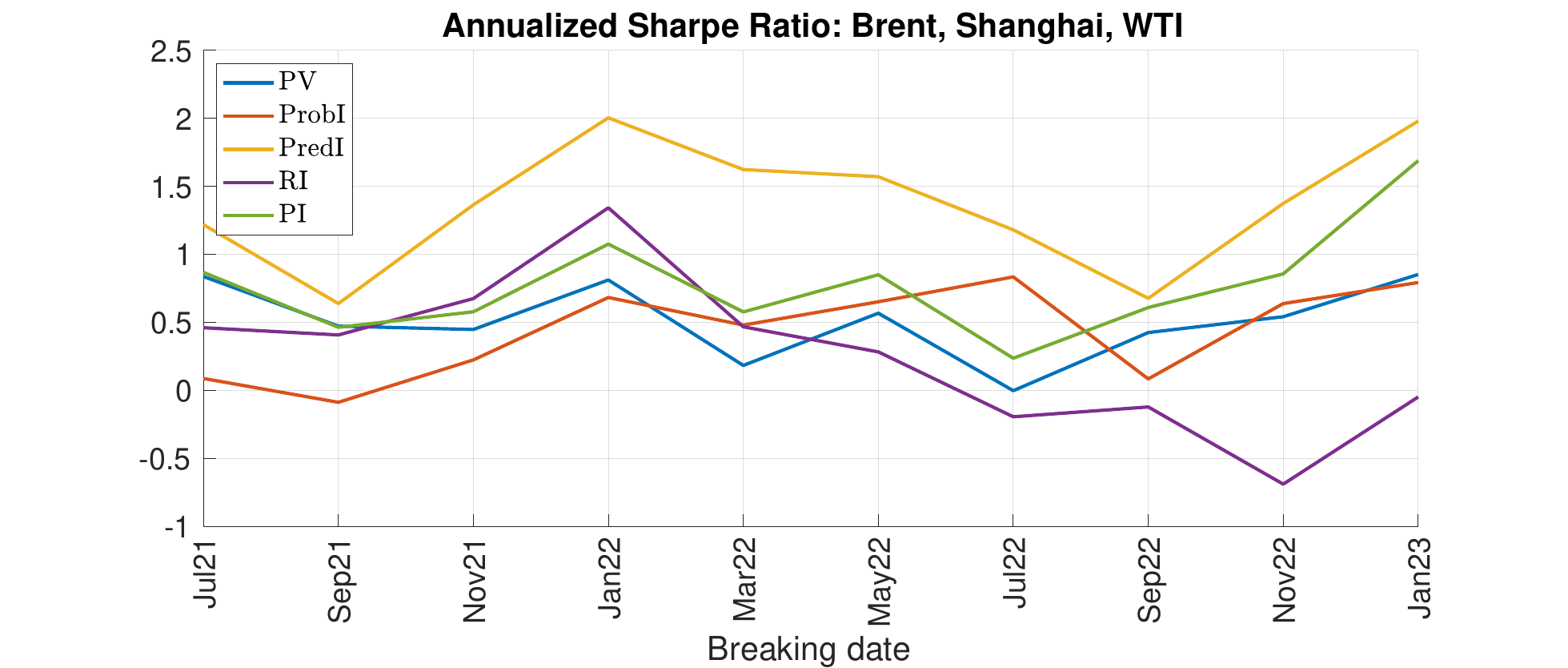}
    \caption{Performances of the strategies with respect to $t_B$; $t_0=$ 03/26/2018.}
    \label{fig:Performances_BSW_tB}
\end{figure}
Figure \ref{fig:Performances_BSW_tB} shows the performance of our strategies as a function of the breaking date $t_B$. As $t_B$ moves backward, the training sample shrinks, while the test sample becomes longer. The impact on the performance of statistical arbitrage strategies is therefore twofold. On the one hand, reducing the size of the training sample weakens the reliability of estimated cointegration model. On the other hand, a larger test sample provides more opportunities for profitable trades. In our dataset, a small decline in the performance is registered for $t_B$ after January 2022. Bringing $t_B$ forward has ambiguous effects as well: the cointegration is estimated over a larger training sample but there is less time to take advantage of that. As shown in Figure \ref{fig:Performances_BSW_tB}, in our dataset we document a small increase of the performances  for $t_B$ after July 2022.

\section{Conclusions}\label{sec:Conclusions}

In this work, we have investigated statistical arbitrage strategies that extend classical pairs trading strategies, involving the two established crude oil futures contracts, the Brent and the WTI, together with the more recently introduced Shanghai crude oil futures. We have documented that the time series of the three futures prices are cointegrated. In order to dynamically model the resulting cointegration spread, we have proposed a mean-reverting process with regime switching modulated by an unobservable Markov chain. The model is estimated by relying on a filter-based version of the EM algorithm, thereby ensuring that the model stays dynamically tuned to the market situation. 
We evaluate several statistical arbitrage strategies based on both backward-looking and model-based forward-looking signals. The empirical results show that strategies grounded in the proposed regime-switching model deliver economically meaningful and robust performance, even under conservative levels of transaction costs. In particular, incorporating the Shanghai crude oil futures enhances profitability, suggesting that this relatively new asset yields relative-value trading opportunities. Moreover, the use of more than two futures contracts enables effective arbitrage execution even when pairwise cointegration is absent.

Based on these results, we believe that it would be interesting to evaluate alternative investment strategies based on our stochastic model and study optimal portfolio problems by relying on the techniques of stochastic optimal control under partial information.

\bibliographystyle{apalike}

\begin{thebibliography}{}

\bibitem[Alizadeh and Nomikos, 2008]{AN08}
Alizadeh, A. and Nomikos, N. (2008).
\newblock Performance of statistical arbitrage in petroleum futures markets.
\newblock {\em Journal of Energy Markets}, 1(2):3--33.

\bibitem[Avellaneda and Lee, 2010]{AL10}
Avellaneda, M. and Lee, J.-H. (2010).
\newblock Statistical arbitrage in the {US} equities market.
\newblock {\em Quantitative Finance}, 10(7):761--782.

\bibitem[Baviera and Baldi, 2019]{BB19}
Baviera, R. and Baldi, T.~S. (2019).
\newblock Stop-loss and leverage in optimal statistical arbitrage with an
  application to energy market.
\newblock {\em Energy Economics}, 79:130--143.

\bibitem[Bertram, 2010]{Ber10}
Bertram, W. (2010).
\newblock Analytic solutions for optimal statistical arbitrage trading.
\newblock {\em Physica A}, 389(11):2234--2243.

\bibitem[Bock and Mestel, 2009]{BM09}
Bock, M. and Mestel, R. (2009).
\newblock A regime-switching relative value arbitrage rule.
\newblock In Fleischmann, B., Borgwardt, K.-H., Klein, R., and Tuma, A.,
  editors, {\em Operations Research Proceedings 2008}, pages 9--14, Berlin,
  Heidelberg. Springer.

\bibitem[Bondarenko, 2003]{Bon03}
Bondarenko, O. (2003).
\newblock Statistical arbitrage and securities prices.
\newblock {\em Review of Financial Studies}, 16(3):875--919.

\bibitem[Burgess, 1999]{Burgess}
Burgess, A.~N. (1999).
\newblock {\em A Computational Methodology for Modelling the Dynamics of
  Statistical Arbitrage}.
\newblock PhD thesis, London Business School.

\bibitem[Caporin et~al., 2019]{CFT19}
Caporin, M., Fontini, F., and Talebbeydokhti, E. (2019).
\newblock Testing persistence of {WTI} and {B}rent long-run relationship after
  the shale oil supply shock.
\newblock {\em Energy Economics}, 79:21--31.

\bibitem[Cerqueti and Fanelli, 2021]{cerqueti2021long}
Cerqueti, R. and Fanelli, V. (2021).
\newblock Long memory and crude oil’s price predictability.
\newblock {\em Annals of Operations Research}, 299:895--906.

\bibitem[Cerqueti et~al., 2019]{cerqueti2019long}
Cerqueti, R., Fanelli, V., and Rotundo, G. (2019).
\newblock Long run analysis of crude oil portfolios.
\newblock {\em Energy Economics}, 79:183--205.

\bibitem[Cotter et~al., 2022]{cotter2022commodity}
Cotter, J., Eyiah-Donkor, E., and Pot{\`\i}, V. (2022).
\newblock Commodity futures return predictability and intertemporal asset
  pricing.
\newblock {\em Journal of Commodity Markets}, page 100289.

\bibitem[Crépellière et~al., 2023]{CrepellierePelsterZeisberger2023}
Crépellière, T., Pelster, M., and Zeisberger, S. (2023).
\newblock Arbitrage in the market for cryptocurrencies.
\newblock {\em Journal of Financial Markets}, 64:100817.

\bibitem[Cummins and Bucca, 2012]{CB12}
Cummins, M. and Bucca, A. (2012).
\newblock Quantitative spread trading on crude oil and refined products
  markets.
\newblock {\em Quantitative Finance}, 12(12):1857--1875.

\bibitem[Do and Faff, 2010]{DF10}
Do, B. and Faff, R. (2010).
\newblock Does simple pairs trading still work?
\newblock {\em Financial Analysts Journal}, 66(4):83--95.

\bibitem[Do and Faff, 2012]{DF12}
Do, B. and Faff, R. (2012).
\newblock Are pairs trading profits robust to trading costs?
\newblock {\em Journal of Financial Research}, 35(2):261--287.

\bibitem[Dunis et~al., 2006]{DLE06}
Dunis, C., Laws, J., and Evans, B. (2006).
\newblock Trading futures spreads: an application of correlation and threshold
  filters.
\newblock {\em Applied Financial Economics}, 16(12):903--914.

\bibitem[Elliott et~al., 1995]{EAM}
Elliott, R.~J., Aggoun, L., and Moore, J.~B. (1995).
\newblock {\em Hidden Markov Models: Estimation and Control}.
\newblock Springer, New York.

\bibitem[Elliott and Bradrania, 2018]{EB18}
Elliott, R.~J. and Bradrania, R. (2018).
\newblock Estimating a regime switching pairs trading model.
\newblock {\em Quantitative Finance}, 18(5):877--883.

\bibitem[Elliott et~al., 2005]{EHM05}
Elliott, R.~J., {Van der Hoek}, J., and Malcolm, W.~P. (2005).
\newblock Pairs trading.
\newblock {\em Quantitative Finance}, 5(3):271--276.

\bibitem[Endres and Stübinger, 2019]{ES19}
Endres, S. and Stübinger, J. (2019).
\newblock A flexible regime switching model with pairs trading application to
  the s\&p 500 high-frequency stock returns.
\newblock {\em Quantitative Finance}, 19(10):1727--1740.

\bibitem[Engle and Granger, 1987]{EG87}
Engle, R. and Granger, C. (1987).
\newblock Co-integration and error correction: representation, estimation and
  testing.
\newblock {\em Econometrica}, 55:251--276.

\bibitem[Erlwein et~al., 2010]{EBM10}
Erlwein, C., Benth, F.~E., and Mamon, R. (2010).
\newblock {HMM} filtering and parameter estimation of an electricity spot price
  model.
\newblock {\em Energy Economics}, 32(5):1034--1043.

\bibitem[Erlwein and Mamon, 2009]{EM09}
Erlwein, C. and Mamon, R. (2009).
\newblock An online estimation scheme for a {Hull-White} model with
  {HMM}-driven parameters.
\newblock {\em Statistical Methods and Applications}, 18:87--107.

\bibitem[Focardi et~al., 2016]{FocardiFabozziMitov2016}
Focardi, S.~M., Fabozzi, F.~J., and Mitov, I.~K. (2016).
\newblock A new approach to statistical arbitrage: Strategies based on dynamic
  factor models of prices and their performance.
\newblock {\em Journal of Banking and Finance}, 65:134--155.

\bibitem[Fontana and Runggaldier, 2010]{FR10}
Fontana, C. and Runggaldier, W.~J. (2010).
\newblock Credit risk and incomplete information: filtering and {EM} parameter
  estimation.
\newblock {\em International Journal of Theoretical and Applied Finance},
  13(5):683--715.

\bibitem[Galay, 2019]{G19}
Galay, G. (2019).
\newblock Are crude oil markets cointegrated? {T}esting the co-movement of
  weekly crude oil spot prices.
\newblock {\em Journal of Commodity Markets}, 16:100088.

\bibitem[Gatev et~al., 2006]{GGR06}
Gatev, E., Goetzmann, W.~N., and Rouwenhorst, K.~G. (2006).
\newblock Pairs trading: performance of a relative-value arbitrage rule.
\newblock {\em Review of Financial Studies}, 19(3):797--827.

\bibitem[Grimm et~al., 2020]{GEM20}
Grimm, S., Erlwein-Sayer, C., and Mamon, R. (2020).
\newblock Discrete-time implementation of continuous-time filters with
  application to regime-switching dynamics estimation.
\newblock {\em Nonlinear Analysis: Hybrid Systems}, 35:100814.

\bibitem[Guidolin and Pedio, 2018]{GuidolinPedio}
Guidolin, M. and Pedio, M. (2018).
\newblock {\em Essentials of time series for financial applications}.
\newblock Academic Press, San Diego (CA).

\bibitem[Hain et~al., 2018]{HainHessUhrigHomburg2018}
Hain, M., Hess, J., and Uhrig-Homburg, M. (2018).
\newblock Relative value arbitrage in {E}uropean commodity markets.
\newblock {\em Energy Economics}, 69:140--154.

\bibitem[Hamilton, 1994]{Hamilton}
Hamilton, J.~D. (1994).
\newblock {\em Time Series Analysis}.
\newblock Princeton University Press, Princeton (NJ).

\bibitem[Hammoudeh et~al., 2008]{HET08}
Hammoudeh, S., Ewing, B., and Thompson, M. (2008).
\newblock Threshold cointegration analysis of crude oil benchmarks.
\newblock {\em The Energy Journal}, 29(4):79--95.

\bibitem[Huang and Huang, 2020]{HH20}
Huang, X. and Huang, S. (2020).
\newblock Identifying the comovement of price between {C}hina{`}s and
  international crude oil futures: a time-frequency perspective.
\newblock {\em International Review of Financial Analysis}, 72:101562.

\bibitem[Ingersoll, 1987]{Ingersoll}
Ingersoll, J. (1987).
\newblock {\em Theory of Financial Decision Making}.
\newblock Rowman \& Littlefield, Savage (MD).

\bibitem[Johansen, 1988]{J88}
Johansen, S. (1988).
\newblock Statistical analysis of cointegration vectors.
\newblock {\em Journal of Economic Dynamics and Control}, 12(2):231--254.

\bibitem[Johansen, 1995]{Johansen}
Johansen, S. (1995).
\newblock {\em Likelihood-based inference in cointegrated vector autoregressive
  models}.
\newblock Okford University Press (UK).

\bibitem[Kristoufek and Vosvrda, 2014]{KV14}
Kristoufek, L. and Vosvrda, M. (2014).
\newblock Commodity futures and market efficency.
\newblock {\em Energy Econmics}, 42:50--57.

\bibitem[Kwiatkowski et~al., 1992]{KPSS92}
Kwiatkowski, D., Phillips, P., Schmidt, P., and Shin, Y. (1992).
\newblock Testing the null hypothesis of stationarity against the alternative
  of a unit root: How sure are we that economic time series have a unit root?
\newblock {\em Journal of Econometrics}, 54(1):159--178.

\bibitem[Ledoit and Wolf, 2008]{LedoitWolf2008}
Ledoit, O. and Wolf, M. (2008).
\newblock Robust performance hypothesis testing with the {S}harpe ratio.
\newblock {\em Journal of Empirical Finance}, 15(5):850--859.

\bibitem[Lee and Papanicolaou, 2016]{LP16}
Lee, S. and Papanicolaou, A. (2016).
\newblock Pairs trading of two assets with uncertainty in co-integration's
  level of mean reversion.
\newblock {\em International Journal of Theoretical and Applied Finance},
  19(8):1650054.

\bibitem[Marshall et~al., 2013]{MarshallNguyenVisaltanachoti2013}
Marshall, B.~R., Nguyen, N.~H., and Visaltanachoti, N. (2013).
\newblock {ETF} arbitrage: Intraday evidence.
\newblock {\em Journal of Banking and Finance}, 37(9):3486--3498.

\bibitem[McLachlan and Krishnan, 2008]{McLachlanKrishnan}
McLachlan, G.~J. and Krishnan, T. (2008).
\newblock {\em The EM Algorithm and its Extensions}.
\newblock Wiley, Hoboken (NJ).

\bibitem[Niu et~al., 2023]{NMC2023}
Niu, J., Ma, C., and Chang, C. (2023).
\newblock The arbitrage strategy in the crude oil futures market of {S}hanghai
  international energy exchange.
\newblock {\em Economic Change and Restructuring}, (56):1201--1223.

\bibitem[Perron, 1997]{P97}
Perron, P. (1997).
\newblock Further evidence on breaking trend functions in macroeconomic
  variables.
\newblock {\em Journal of Econometrics}, 80(2):355--385.

\bibitem[Phillips and Perron, 1988]{PP88}
Phillips, P. and Perron, P. (1988).
\newblock Testing for a unit root in time series regression.
\newblock {\em Biometrika}, 75(2):335--346.

\bibitem[Politis and Romano, 1994]{PolitisRomano1994}
Politis, D.~N. and Romano, J.~P. (1994).
\newblock The stationary bootstrap.
\newblock {\em Journal of the American Statistical Association},
  89(428):1303--1313.

\bibitem[Rein et~al., 2021]{RRS2021}
Rein, C., R{\"u}schendorf, L., and Schmidt, T. (2021).
\newblock Generalized statistical arbitrage concepts and related gain
  strategies.
\newblock {\em Mathematical Finance}, 31(2):563--594.

\bibitem[Said and Dickey, 1984]{SD84}
Said, S. and Dickey, D. (1984).
\newblock Testing for unit roots in autoregressive-moving average models of
  unknown order.
\newblock {\em Biometrika}, 71(3):599--607.

\bibitem[Sarmento and Horta, 2021]{SarmentoHorta}
Sarmento, S. and Horta, N. (2021).
\newblock {\em A {M}achine {L}earning {B}ased {P}airs {T}rading investment
  strategy}.
\newblock Springer, Princeton (NJ).

\bibitem[Sullivan et~al., 1999]{SullivanTimmermannWhite1999}
Sullivan, R., Timmermann, A., and White, H. (1999).
\newblock Data-snooping, technical trading rule performance, and the bootstrap.
\newblock {\em The Journal of Finance}, 54(5):1647--1691.

\bibitem[Tenyakov and Mamon, 2017]{TM17}
Tenyakov, A. and Mamon, R. (2017).
\newblock A computing platform for pairs-trading online implementation via a
  blended {K}alman-{HMM} filtering approach.
\newblock {\em Journal of Big Data}, 4:46.

\bibitem[Vidyamurthy, 2004]{Vidya}
Vidyamurthy, G. (2004).
\newblock {\em Pairs Trading: Quantitative Methods and Analysis}.
\newblock Wiley, Hoboken (NJ).

\bibitem[White, 2000]{White2000}
White, H. (2000).
\newblock A reality check for data snooping.
\newblock {\em Econometrica}, 68(5):1097--1126.

\bibitem[Yang et~al., 2020]{yang2020pricing}
Yang, C., Lv, F., Fang, L., and Shang, X. (2020).
\newblock The pricing efficiency of crude oil futures in the {S}hanghai
  international exchange.
\newblock {\em Finance Research Letters}, 36:101329.

\bibitem[Yang and Zhou, 2020]{YZ20}
Yang, J. and Zhou, Y. (2020).
\newblock Return and volatility transmission between {C}hina{`}s and
  international crude oil futures markets: a first look.
\newblock {\em Journal of Futures Markets}, 40(6):860--884.

\bibitem[Yun and He, 2022]{XH22}
Yun, X. and He, J. (2022).
\newblock Pairs trading and asset pricing.
\newblock {\em Pacific-Basin Finance Journal}, 72:101713.

\end{thebibliography}

\end{document}